\documentclass[a4paper,twocolumn,11pt,accepted=2025-08-12]{quantumarticle}
\pdfoutput=1
\usepackage[utf8]{inputenc}
\usepackage[english]{babel}
\usepackage[T1]{fontenc}
\usepackage{amsmath}
\usepackage{amssymb}
\usepackage{graphicx}
\usepackage{esint}
\usepackage{epstopdf}
\usepackage{braket}
\usepackage{hyperref}
\usepackage{tikz}
\usepackage{lipsum}
\usepackage{stackengine}

\usepackage[numbers,sort&compress]{natbib}

\begin{document}

\title{Simultaneous symmetry breaking in spontaneous Floquet states: temporal Floquet-Nambu-Goldstone modes, Floquet thermodynamics, and the time operator}

\author{Juan Ram\'on Mu\~noz de Nova}
\email{jrmnova@fis.ucm.es}
\affiliation{Departamento de F\'isica de Materiales, Universidad Complutense de Madrid, E-28040 Madrid, Spain}
\orcid{0000-0001-6229-9640}

\author{Fernando Sols}
\email{f.sols@ucm.es}
\affiliation{Departamento de F\'isica de Materiales, Universidad Complutense de Madrid, E-28040 Madrid, Spain}
\orcid{0000-0002-0947-286X}

\maketitle
\begin{abstract}
We study simultaneous symmetry breaking in spontaneous Floquet states, focusing on the specific case of an atomic condensate. We first describe the quantization of the Nambu-Goldstone (NG) modes for a stationary state simultaneously breaking several symmetries of the Hamiltonian by invoking the generalized Gibbs ensemble, which enables a thermodynamical description of the problem. The quantization procedure involves a Berry-Gibbs connection, which depends on the macroscopic conserved charges associated to each broken symmetry and whose curvature is not invariant under generalized gauge transformations. We extend the formalism to Floquet states, where Goldstone theorem translates into the emergence of Floquet-Nambu-Goldstone (FNG) modes with zero quasi-energy. In the case of a spontaneous Floquet state, there is a genuine temporal FNG mode arising from the continuous time-translation symmetry breaking, whose quantum amplitude provides a rare realization of a time operator in Quantum Mechanics. Furthermore, since they conserve energy, spontaneous Floquet states can be shown to possess a conserved Floquet charge. Both the temporal FNG mode and the Floquet charge are distinctive features of spontaneous Floquet states, absent in conventional, driven systems. Nevertheless, as these operate at fixed frequency, they also admit a thermodynamic description in terms of the Floquet enthalpy, the Legendre transform of the energy with respect to the Floquet charge. An important conclusion of our analysis is that certain time crystals may not represent a genuine spontaneous temporal symmetry-breaking, since their periodic behavior just results from constant motion along a closed orbit spawned by the generators of other broken symmetries. We apply our formalism to a particular realization of spontaneous Floquet state, the so-called state of continuous emission of solitons (CES), which breaks $U(1)$ and time-translation symmetries, representing a time supersolid. We numerically compute its density-density correlations, predicted to be dominated by the temporal FNG mode at long times, observing a remarkable agreement between simulation and theory. Based on these results, we propose a feasible experimental scheme to observe the temporal FNG mode of the CES state.
\end{abstract}

\section{Introduction}

Noether and Goldstone theorems are two of the most fundamental results in Physics, and can be regarded as the sides of the same coin \cite{Callen1985}. The former states that any continuous symmetry is translated into a conservation law, while the latter asserts that solutions spontaneously breaking one of those symmetries have associated a zero-energy mode, known as Nambu-Goldstone (NG) mode. Spontaneous symmetry breaking is still a hot topic of research, motivating the search for novel phases of quantum matter which simultaneously break several continuous symmetries, such as supersolids \cite{Leonard2017,Li2017}, quantum Hall ferromagnets \cite{Ezawa2008,Kharitonov2012PRL,Kim2021,Coissard2022}, spinor \cite{stenger1998,Stamper2013,Blinova2023} and rotating \cite{Polkinghorne2021} condensates, or quantum droplets \cite{Chomaz2016,Cabrera2018,Liu2019,Zin2021}.

A fascinating perspective is provided by time crystals \cite{Wilczek2012,Sacha2020}, which spontaneously break time-translation symmetry, perhaps the most fundamental symmetry in nature. Although a no-go theorem ruling out them as first conceived was proven \cite{Watanabe2015}, time crystals are still possible in the out-of-equilibrium arena. Discrete \cite{Choi2017,Zhang2017,Smits2018,Rovny2018,Kyprianidis2021,Randall2021,Google2022,Frey2022} and continuous time crystals \cite{Autti2018,Kongkhambut2022,Dreon2022,Liu2023,Greilich2024} have been observed in a wide variety of systems, ranging from cold atoms to superconducting quantum computers. In particular, discrete time crystals emerge in Floquet systems as a subharmonic response to the external periodic driving \cite{Sacha2015,Else2016}. In general, Floquet systems \cite{Shirley1965,Sambe1973,Grifoni1998} provide rich scenarios to study a number of out-of-equilibrium features such as prethermalization \cite{Peng2021}, topological insulation \cite{Lindner2011}, dynamical phase transitions \cite{Prosen2011}, high-harmonic generation \cite{Murakami2018}, or protected cat states \cite{Pieplow2019} and flat-band superfluidity \cite{Mateos2023}. 

It was recently shown that Floquet physics can also arise in time-independent configurations, where the system self-consistently oscillates as a Floquet state of its effective Hamiltonian due to many-body interactions \cite{deNova2022}. A specific realization of spontaneous Floquet state was proposed using an atomic Bose-Einstein condensate (BEC), the so-called state of continuous emission of solitons (CES) \cite{deNova2016,deNova2021}. As a result, the CES state breaks $U(1)$ and continuous time-translation symmetries, providing the temporal analogue of a supersolid, which has been labeled as time supersolid \cite{Autti2018}. Some natural questions then arise: how to combine several broken symmetries when one of them is time itself? What is the temporal version of Goldstone theorem?

In this work, we study simultaneous symmetry breaking (here defined as the spontaneous symmetry breaking of several continuous symmetries by the same state) in spontaneous Floquet states, focusing on the specific case of an atomic BEC close to zero temperature for illustrative purposes. We first discuss simultaneous symmetry breaking in stationary states within the framework of the generalized Gibbs ensemble \cite{Cazalilla2006,Rigol2007,Langen2015}, since its equitable treatment of each broken symmetry allows for a general and complete description of the problem, including its thermodynamics. The generalized Gibbs ensemble emerges in the study of thermalization in isolated systems, where each conserved charge further restricts the available phase-space for the dynamics, leading to a violation of the eigenstate thermalization hypothesis \cite{Deutsch1991,Srednicki1994,Rigol2008}, in parallel to other mechanisms such as many-body localization \cite{Basko2006,Schreiber2015} or Hilbert space fragmentation \cite{Sala2020}.

The NG modes appear as zero-energy modes of the spectrum of collective excitations. Each NG mode is paired with a Gibbs mode, which accounts for the fluctuations of the quantum state with respect to the associated conserved charge, stemming from the fact that a symmetry-broken state cannot be an eigenstate of the charge generating the corresponding symmetry transformation. Remarkably, the Goldstone-Gibbs modes can be described by a geometrical approach that involves a generalization of the Berry connection, defined within a manifold whose variables are the conserved charges, denoted as a result as the Berry-Gibbs connection. This connection is extended to the full manifold which includes the continuous parameters of the broken symmetry transformations, where its curvature gives rise to a symplectic form that provides the commutation relations between the amplitudes of the Goldstone-Gibbs modes, hence behaving as conjugate coordinate-momenta. The dynamics inherits this correspondence, where the amplitudes of the Gibbs modes behave as  conserved momenta, since they represent the quantum fluctuations of the corresponding  charges, while the amplitudes of the Goldstone modes display a linear time dependence, with a velocity linear in the momenta. Our results for simultaneous symmetry breaking in a stationary state thus generalize the original works of Refs. \cite{Lewenstein1996,Dziarmaga2004} to an arbitrary number of broken symmetries, and provide a deeper understanding of the nature of the Goldstone-Gibbs modes.

As an example, we apply our formalism to a cnoidal wave in a superfluid, a model that has been used to understand fundamental aspects of supersolidity \cite{Martone2021}, validating our theoretical results. An important conclusion of this study is that periodic motion along a closed orbit spawned by the broken-symmetry generators may result in the misidentification of time-crystalline behavior, not representing a genuine symmetry-breaking of time-translation invariance.

The generality of the developed framework allows its application for the study of simultaneous symmetry breaking in Floquet states, where the NG modes are now translated into Floquet-Nambu-Goldstone (FNG) modes with zero quasi-energy. In particular, since spontaneous Floquet states break continuous time-translation symmetry, they present a genuine temporal FNG mode. For a complete characterization of the problem, we develop the $(t,\phi)$ formalism, denoted in this way because of its analogy with the standard $(t,t')$ formalism \cite{Peskin1993,Heimsoth2012}. In combination with the generalized Gibbs ensemble, the $(t,\phi)$ formalism leads to a thermodynamic description analogous to that of stationary states, which we term as Floquet thermodynamics. For spontaneous Floquet states, this is a direct consequence of energy conservation, which implies the existence of a conserved Floquet charge. Conventional Floquet states also admit a thermodynamic description in terms of the Floquet enthalpy, the Legendre transform of the energy with respect to the Floquet charge. This is because driven systems are isoperiodic, i.e., they operate at an externally fixed frequency, which is the conjugate variable of the Floquet charge, in analogy with isobaric or isothermal systems where the environment fixes the pressure or the temperature, respectively. The temporal FNG mode and the Floquet charge are thus the distinctive features of a spontaneous Floquet state, absent in driven systems. 

The quantization of the Goldstone-Gibbs modes is performed along the same lines as in the stationary case, described by a Berry-Gibbs connection that results in a similar coordinate-momentum correspondence. However, a novel feature is that the amplitude of the temporal FNG mode can be regarded as an effective time coordinate, whose conjugate momentum represents the energy fluctuations. Therefore, the quantum amplitude of the temporal FNG mode represents a unique realization of a time operator in Quantum Mechanics. 

Finally, we apply our formalism to the CES state, computing its quantum fluctuations with the help of the Truncated Wigner method \cite{Sinatra2002,Carusotto2008}. Theory predicts that, at long times, the temporal FNG mode dominates the density-density correlations, observing an excellent agreement with the numerical results. Based on these results, we propose a feasible experimental scheme to observe the temporal FNG mode of the CES state.

The article is arranged as follows. Section \ref{sec:GoldstoneGeneral} discusses how the general results from Goldstone theorem are recovered in variational approaches in many-body systems. Section \ref{sec:SimultaneousSymmetryBreaking} addresses simultaneous symmetry breaking in stationary states within the generalized Gibbs ensemble, giving rise to the concept of Berry-Gibbs connection. Section \ref{sec:Cnoidal} applies the formalism of Sec. \ref{sec:SimultaneousSymmetryBreaking} to a cnoidal wave in a superfluid. Section \ref{sec:GoldstoneFloquetGeneral} extends the results of Sec. \ref{sec:GoldstoneGeneral} to Floquet states, both spontaneous and conventional. Section \ref{sec:SimultaneousSymmetryBreakingFloquet} discusses simultaneous symmetry breaking in Floquet systems using the $(t,\phi)$ formalism within the generalized Gibbs ensemble. This leads to the emergence of Floquet thermodynamics, Sec. \ref{subsec:FloquetGPG}, and of a time operator, Sec. \ref{subsec:TimeOperator}. Section \ref{sec:ces} applies the formalism of Sec. \ref{sec:SimultaneousSymmetryBreakingFloquet} to study the quantum fluctuations of the CES state and proposes an experimental setup for the observation of the temporal FNG mode. Technical details are presented in the Appendix.

\section{Variational Goldstone Theorem}\label{sec:GoldstoneGeneral}

Originally derived within the framework of relativistic quantum field theory \cite{Goldstone1962}, the general idea behind Goldstone theorem is universal, namely, solutions which break some continuous symmetry of the equations of motion include a linear zero mode associated to the broken symmetry. This feature is even present in classical mechanics. Consider a particle of mass $m$ moving within a potential $V(\mathbf{x})$ which is invariant under some continuous symmetry transformation of the coordinates, $V(\mathbf{x}')=V(\mathbf{x})$, $\mathbf{x}'=\mathbf{x}'(\mathbf{x},\alpha)$, where $\alpha$ parameterizes the transformation in such a way that $\alpha=0$ corresponds to the identity transformation, $\mathbf{x}'(\mathbf{x},0)=\mathbf{x}$. As standard in Lie theory, the symmetry transformation can be expanded as
\begin{equation}\label{eq:LieTransformation}
    \mathbf{x}'=\mathbf{x}+\alpha \boldsymbol{\xi}+\ldots
\end{equation}
where the vector $\boldsymbol{\xi}=-iL\mathbf{x}$ describes the symmetry transformation at the infinitesimal level and $L$ is the generator of the continuous transformation on the coordinates. At the same time, Noether theorem guarantees the existence of a conserved quantity associated to that symmetry.  

Time-independent solutions to the equations of motion are given by the equilibrium points $\mathbf{x}_0$ which extremize the potential, $\boldsymbol{\nabla}V(\mathbf{x}_0)=0$. Linearized motion around equilibrium positions is characterized by the eigenvalues $m\Omega^2_i$ of the Hessian matrix $\partial^2_{ij}V(\mathbf{x}_0)$. In particular, potential minima are stable equilibrium points, as then all $\Omega_i$ are real and represent the frequencies of the normal modes of oscillation. 

It is immediate to show that the symmetry of the potential implies
\begin{equation} \partial^2_{ij}V(\mathbf{x}_0)\xi_0^j=0,~\boldsymbol{\xi}_0=-iL\mathbf{x}_0,
\end{equation}
where hereafter we use Einstein convention summation unless otherwise stated. Thus, if the symmetry is broken by the equilibrium position, which means that it is not invariant under the symmetry transformation, $L\mathbf{x}_0\neq 0$, then there is a zero-frequency normal mode, the Nambu-Goldstone mode. Qualitatively, this zero mode results from the fact that there is no restoring force along the orbit of the equilibrium point under the symmetry transformation because potential energy is conserved there.

A simple example to visualize this behavior is given by a particle in a 2D $y$-independent potential, $V(x,y)=V(x)$, with $V(0)$ a global minimum and $V''(0)= m\omega_x^2$. In that case, if we consider that the particle is initially placed at the equilibrium position $\mathbf{x}=(0,y_0)$, small displacements along the $x$-axis lead to oscillations with frequency $\omega_x$. However, due to the symmetry of the problem, the momentum along the $y$-axis $p_y$ is conserved and any small displacement along this direction leads to an unbounded motion with constant velocity $v_y=p_y/m$, $y(t)=y_0+v_yt$, since there is no restoring force in this direction.

The previous concepts can be straightforwardly extended to quantum many-body systems. In that context, it is common to use variational approaches based on trial wavefunctions characterized by a reduced number of parameters in order to tackle the exponentially large complexity of the problem. For example, in the Dirac-Frenkel variational principle, the \textit{ansatz} state $\ket{\Psi(t)}$ is required to extremize 
\begin{equation}\label{eq:DiracFrenkel}
    L(t)\equiv\bra{\Psi(t)}i\hbar\partial_t-\hat{H}\ket{\Psi(t)},
\end{equation}
which would yield the exact Schr\"odinger equation if no restriction were imposed on $\ket{\Psi(t)}$\footnote{Technically, the extremization condition is $\bra{\delta\Psi(t)} i\hbar\partial_t-\hat{H}\ket{\Psi(t)}=0$.}. The Dirac-Frenkel formalism can also be extended to finite temperatures \cite{DeAngelis1991}. Other variational methods are based on a Lagrangian approach, where the trial wavefunction extremizes some action; indeed, Eq. (\ref{eq:DiracFrenkel}) can be regarded as an effective Lagrangian. A large number of well-known equations fall under this class of approximation, including the Gross-Pitaevskii (GP) equation \cite{Pitaevskii2016}, the Hartree-Fock (HF) equations \cite{Giuliani2005}, the Gutzwiller ansatz \cite{Jaksch1998}, or the MultiConfiguration Time-Dependent Hartree (MCTDH) method \cite{Caillat2005,Alon2008}. In all cases, if we group the variational parameters in a generic vector $\mathbf{X}$, the resulting equations of motion take the form
\begin{equation}\label{eq:SelfConsistentVariational}
    i\hbar\frac{dX^i}{dt}=H^i_{j}(\mathbf{X})X^j,
\end{equation} 
where the vector components $X^i$ can be orbital wavefunctions (GP and HF equations), wave-function coefficients (Gutzwiller ansatz), or even both (MCTDH method). The matrix $H^i_{j}(\mathbf{X})$ usually depends on the vector $\mathbf{X}$ due to many-body interactions, acting as the effective Hamiltonian of the problem. For the present moment, we take the original Hamiltonian $\hat{H}$ as time-independent, so $H^i_{j}(\mathbf{X})$ only depends implicitly on time through $\mathbf{X}$. 

Now, we assume that there are stationary solutions of the form $X^i(t)=X_0^ie^{-i\epsilon^i t/\hbar}$, satisfying the self-consistent eigenvalue equation
\begin{equation}\label{eq:SelfConsistentVariationalStationary}
    \epsilon^{i}X_0^i=H^i_{j}(\mathbf{X}_0) X_0^j.
\end{equation}
The spectrum of collective modes associated to this stationary solution is obtained by considering small perturbations around  $\mathbf{X}_0$, $X^i(t)=[X_0^i+\delta X^i(t)]e^{-i\epsilon^i t/\hbar}$, which yields a linear equation of the general form
\begin{equation}\label{eq:VariationalCollectiveModes}
    i\hbar\frac{d\delta X^i}{dt}=M^i_{j}(\mathbf{X}_0)\delta X^{j},
\end{equation}
with $M^i_{j}(\mathbf{X}_0)$ some matrix that depends on the background stationary solution. The eigenvalues of $M^i_{j}(\mathbf{X}_0)$ are typically the energies of the collective modes. This is indeed the case of the Bogoliubov-de Gennes (BdG) equations for both the GP equation \cite{Pitaevskii2016} and the Gutzwiller ansatz \cite{Caleffi2020}, and of the Time-Dependent Hartree-Fock Approximation (TDHFA) for the HF equations \cite{Thouless1961}.

In the same fashion as in the classical example, if the stationary solution $\mathbf{X}_0$ spontaneously breaks a continuous symmetry of the problem, so $\mathbf{X}_0'=\mathbf{X}_0'(\mathbf{X}_0,\alpha)=\mathbf{X}_0-i\alpha L \mathbf{X}_0+\ldots$ is also a stationary solution, then the vector $\mathbf{Y}\equiv-i L\mathbf{X}_0\neq 0$ is a zero mode of the linearized equations of motion,
\begin{equation}\label{eq:VariationalGoldstone}
    0=M^i_{j}(\mathbf{X}_0)Y^j.
\end{equation}
This is the NG mode associated to the spontaneous symmetry breaking. We note that this derivation also applies to pseudo-NG modes \cite{Weinberg1972,Georgi1975}, which emerge from extended symmetries in the equations of motion that were not present in the original Hamiltonian. It can be also applied to few-body or open quantum systems governed by similar self-consistent dynamics.

\section{Simultaneous symmetry breaking in stationary states}\label{sec:SimultaneousSymmetryBreaking}

We proceed to review how symmetries emerge in many-body systems. For the sake of definiteness, we consider the following general time-independent second-quantization Hamiltonian, valid for both interacting bosons and fermions \cite{Fetter2003},
\begin{align}\label{eq:HamiltonianManyBody}
&\hat{H}=\int\mathrm{d}\mathbf{x}~\hat{\Psi}^{\dagger}(\mathbf{x})\left[-\frac{\hbar^2}{2m}\nabla^2+V(\mathbf{x})\right]\hat{\Psi}(\mathbf{x})\\
\nonumber&+\frac{1}{2}\int\mathrm{d}\mathbf{x}\int\mathrm{d}\mathbf{x}'\hat{\Psi}^{\dagger}(\mathbf{x})\hat{\Psi}^{\dagger}(\mathbf{x}')\mathbf{V}(\mathbf{x}-\mathbf{x}')\hat{\Psi}(\mathbf{x}')\hat{\Psi}(\mathbf{x}),
\end{align}
where the field operator $\hat{\Psi}$, the external potential $V(\mathbf{x})$, and the interacting potential $\mathbf{V}(\mathbf{x}-\mathbf{x}')$ may present a tensorial structure due to internal degrees of freedom (spin, pseudospin\ldots). The corresponding Heisenberg equation of motion for the field operator $\hat{\Psi}(\mathbf{x},t)=e^{i\hat{H}t/\hbar}\hat{\Psi}(\mathbf{x})e^{-i\hat{H}t/\hbar}$ is
\begin{align}\label{eq:HeisenbergEquationOfMotion}
    \nonumber &i\hbar\partial_t\hat{\Psi}(\mathbf{x},t)=[\hat{\Psi}(\mathbf{x},t),\hat{H}]\\
    \nonumber &=\left[-\frac{\hbar^2}{2m}\nabla^2+V(\mathbf{x})\right]\hat{\Psi}(\mathbf{x},t)\\ &+\int\mathrm{d}\mathbf{x}'~\hat{\Psi}^{\dagger}(\mathbf{x}',t)\mathbf{V}(\mathbf{x}-\mathbf{x}')\hat{\Psi}(\mathbf{x}',t)\hat{\Psi}(\mathbf{x},t).
\end{align}
The classical version of this equation can be derived from the Lagrangian 
\begin{equation}\label{eq:ClassicalLagrangian}
   L\equiv \int\mathrm{d}\mathbf{x}\, \mathcal{L},~\mathcal{L}=i\hbar\Psi^\dagger \partial_t\Psi-\mathcal{H},
\end{equation}
with $\mathcal{H}$ the Hamiltonian density arising from Eq. (\ref{eq:HamiltonianManyBody}). The canonical momentum associated to the field $\Psi$ is $\Pi(\mathbf{x})=i\hbar\Psi^\dagger (\mathbf{x})$, giving rise to the Poisson bracket 
\begin{equation}\label{eq:CanonicalCommutationClassical}
    i\hbar \{\Psi(\mathbf{x}),\Psi^\dagger (\mathbf{x}')\}=\delta(\mathbf{x}-\mathbf{x}'). 
\end{equation}
This Lagrangian correspondence allows, via Noether theorem, to identify the conserved charges arising from the symmetries of the Hamiltonian. Specifically, any continuous symmetry transformation that leaves invariant the action gives rise to a conserved charge
\begin{equation}\label{eq:ClassicalNoetherCharge}
    Q=\int\mathrm{d}\mathbf{x}\left[\xi^0 \mathcal{L}+i\hbar\Psi^\dagger \Delta\Psi    \right].
\end{equation}
In the above equation, $\xi,\Delta \Psi$ describe the infinitesimal variations of the coordinates and the field under the symmetry transformation, respectively:
\begin{eqnarray}
    \nonumber x'^\mu&=&x^\mu+\alpha \xi^\mu(x)+\ldots\\
    \Psi'(x)&=&\Psi(x)+\alpha\Delta \Psi(x)+\ldots
\end{eqnarray}
where we use the standard relativistic notation $x^\mu=(t,\mathbf{x})$. In particular, $\Delta\Psi=-iT\Psi$, with $T$ the generator of the transformation on the field. 




The most common symmetries are invariance under phase transformations, temporal and spatial translations, and rotations, leading to the conservation of particle number $N$, energy $E$ and momentum $\mathbf{P}$, and angular momentum $\mathbf{L}$:
\begin{align}
\Psi'=\Psi e^{-i\theta}&\Longrightarrow N=\int\mathrm{d}\mathbf{x}~\Psi^\dagger\Psi,\\
 \nonumber t'=t-t_0&\Longrightarrow E\equiv H=\int\mathrm{d}\mathbf{x}~\mathcal{H},\\
\nonumber \mathbf{x}'=\mathbf{x}+\mathbf{x}_0&\Longrightarrow \mathbf{P}=\int\mathrm{d}\mathbf{x}~\Psi^\dagger(-i\hbar\boldsymbol{\nabla})\Psi,\\
\nonumber  \mathbf{x}'=  R(\boldsymbol{\varphi})\mathbf{x}&\Longrightarrow\mathbf{L}=\int\mathrm{d}\mathbf{x}~\Psi^\dagger[-i\hbar(\mathbf{x}\times \boldsymbol{\nabla})+\mathbf{S}]\Psi,
\end{align}
where, in the last line, the spin $\mathbf{S}$ arises due to the internal structure of the field $\Psi$.

In the following, lowercase Greek indices $\alpha,\beta\ldots$ label symmetry transformations, $\alpha=\theta, t\ldots$, while uppercase Latin indices $A,B,\ldots$ label the conserved charges $Q_{\alpha,\beta,\ldots}=N,E,\ldots$ associated to the symmetries $\alpha,\beta,\ldots$ The continuous symmetry parameters, the generators and the conserved charges are grouped in vectors $\boldsymbol{\alpha}$, $\mathbf{T}$, $\mathbf{Q}$, respectively. Finally, lowercase Latin indices $a,b,\ldots$ indistinctively label continuous symmetry transformations or the associated conserved charges, $a=\alpha,A$. A summary of the notation is presented in Table \ref{Table:Noether}.


It is well-known that the conserved charge $Q_\alpha$ generates the corresponding symmetry transformation $\alpha$, as revealed by the Poisson bracket
\begin{equation}\label{eq:ChargeGenerationClassical}
    \{\Psi(\mathbf{x}),Q_\alpha\}=\frac{\delta Q_\alpha}{\delta i\hbar \Psi^\dagger}=\Delta \Psi=-iT_\alpha \Psi.
\end{equation}
In particular, most of the symmetries do not involve the time coordinate, $\xi^0=0$, and hence their conserved charges can be written as a simple bilinear expression in the field
\begin{equation}\label{eq:ChargeScalar}
Q_\alpha=\hbar\int\mathrm{d}\mathbf{x}~\Psi^\dagger T_\alpha \Psi=\hbar \braket{\Psi|T_\alpha|\Psi},
\end{equation}
with
\begin{equation}
\braket{\chi|\Psi}\equiv \int\mathrm{d}\mathbf{x}~\chi^\dagger (\mathbf{x})\Psi(\mathbf{x})
\end{equation}
the usual scalar product. The generators $T_\alpha$ are Hermitian operators under this scalar product, so symmetry transformations are implemented by unitary operators of the form $U=e^{-i\boldsymbol{\alpha}\cdot\mathbf{T}}$. We note that these derivations apply for an arbitrary many-body system, including both bosons and fermions.  


\subsection{Spontaneous symmetry breaking in the generalized Gibbs ensemble}\label{subsec:GPG}

We now address the case in which the state of the system spontaneously breaks $n$ continuous symmetries, where in the following we restrict the vectors $\boldsymbol{\alpha},\mathbf{Q},\mathbf{T}$ to have $n$ components corresponding to those broken symmetries. Furthermore, we assume that those symmetries do not involve time, so their charges take the form of Eq. (\ref{eq:ChargeScalar}), and that their generators commute between themselves, 
\begin{equation}
    [T_\alpha,T_\beta]=0.
\end{equation}

In order to explicitly account for spontaneous symmetry breaking in the dynamics, we introduce a Lagrange multiplier $\lambda^\alpha$ for the conserved charge $Q_\alpha$ associated to each broken symmetry,
\begin{equation}\label{eq:LagrangianGibbs} L=\int\mathrm{d}\mathbf{x}~i\Psi^\dagger\partial_t\Psi-H+\lambda^\alpha Q_\alpha.
\end{equation}
By assumption, the broken symmetries do not involve time and commute between themselves, so this constrained Lagrangian yields exactly the same conserved charges $Q_\alpha$. However, the Hamiltonian is now replaced by the generalized Gibbs Hamiltonian
\begin{equation}\label{eq:GibbsHamiltonian}
    K=H-\lambda^\alpha Q_\alpha=H-\boldsymbol{\lambda}\cdot \mathbf{Q}. 
\end{equation}

\begin{table}[!tb]
\centering
\begin{tabular}[c]{|c|c|c|c|}
\hline
~ &  Symmetry & Generator &  Charge\\
\hline
Label &  $\alpha$ & $T_\alpha$ &  $A\equiv Q_\alpha$\\
\hline
Vector &  $\boldsymbol{\alpha}$ & $\mathbf{T}$ &  $\mathbf{Q}$\\
\hline
Example &  $(\theta,t,\mathbf{x})$ & $(1,i\partial_t,-i\boldsymbol{\nabla})$ &  $(N,E,\mathbf{P})$\\
\hline
\end{tabular}
\caption{Summary of notation employed in this work.}
\label{Table:Noether}
\end{table}

For illustrative purposes, we consider the specific case of a BEC of spin-0 particles close to zero temperature, where the classical scalar field $\Psi$ describes the macroscopic wavefunction of the condensate. Hereafter, we take the pseudopotential $\mathbf{V}(\mathbf{x}-\mathbf{x}')=g\delta(\mathbf{x}-\mathbf{x}')$ as the interacting potential, and set units such that $\hbar=m=gn_0=1$, where $n_0$ is some characteristic density that rescales the field so it becomes dimensionless, $\Psi\to \sqrt{n_0}\Psi$. The condensate dynamics is determined by the classical equation of motion derived from the Lagrangian (\ref{eq:ClassicalLagrangian}), which is the usual time-dependent GP equation
\begin{equation}\label{eq:GPEquation}
   i\partial_t\Psi=\left[-\frac{\nabla^2}{2}+V(\mathbf{x})+|\Psi|^2\right]\Psi\equiv H_{GP}\Psi,
\end{equation}
with $H_{GP}$ the effective nonlinear GP Hamiltonian. This equation takes the same self-consistent form of Eq. (\ref{eq:SelfConsistentVariational}) by choosing $\mathbf{X}$ as a two-component vector $\mathbf{X}=[\Psi,\Psi^*]^T$, so then $H(\mathbf{X})$ is a diagonal matrix, $H(\mathbf{X})=\textrm{diag}[H_{GP},-H_{GP}]$. 

Symmetry-breaking states are described as stationary solutions of the constrained Lagrangian (\ref{eq:LagrangianGibbs}), obtained by extremizing $K$ with respect to variations of the macroscopic wavefunction, leading to the Gross-Pitaevskii-Gibbs (GPG) equation:
\begin{equation}\label{eq:GPGEquation}
    \frac{\delta K}{\delta \Psi^*}=0\Longrightarrow K_{GP}\Psi_0\equiv [H_{GP}-\boldsymbol{\lambda}\cdot \mathbf{T}]\Psi_0=0.
\end{equation}
We remark that each component $T_\alpha\Psi_0$ of the vector $\mathbf{T}\Psi_0$ is nonvanishing as they reflect broken symmetries by assumption. Furthermore, we also assume in the following that they are linearly independent, meaning that there is no constant vector $\boldsymbol{u}$ satisfying $u^\alpha T_\alpha \Psi_0=0$. If this were the case, the conserved charges would not be independent between themselves because then $u^\alpha Q_\alpha=0$. Nevertheless, even in this situation, one could further restrict the vectors $\boldsymbol{\alpha},\mathbf{T},\mathbf{Q}$ in such a way that the components of $\mathbf{T}\Psi_0$ are all linearly independent.


The standard time-independent GP equation is retrieved from the GPG equation for $\lambda^\alpha=0,~\alpha\neq \theta$, and $\lambda^\theta=\mu$, where $\mu$ is the chemical potential, the Lagrange multiplier associated to the particle number $N$, emerging due to the spontaneous symmetry breaking of the $U(1)$-invariance under phase transformations. When additional continuous symmetries are spontaneously broken, the generalized Gibbs ensemble allows us to treat all of them on equal grounds, leading to the GPG equation. 


The actual dynamics of the symmetry-breaking state is obtained by inserting the stationary GPG wavefunction into the original time-dependent GP equation (\ref{eq:GPEquation}). Taking as initial condition $\Psi(\mathbf{x},0)=\Psi_0(\mathbf{x})$ gives
\begin{align}\label{eq:TimeDependentGPGEquation}
    &i\partial_t\Psi=H_{GP}\Psi=(\boldsymbol{\lambda}\cdot \mathbf{T}) \Psi\Longrightarrow\\
    \nonumber & \Psi(\mathbf{x},t)=e^{-i\boldsymbol{\lambda}\cdot \mathbf{T} t} \Psi_0(\mathbf{x}).
\end{align}
Hence, the Lagrange multipliers $\boldsymbol{\lambda}$ are the velocities along the orbits generated by the broken-symmetry transformations, $\dot{\boldsymbol{\alpha}}=\boldsymbol{\lambda}$. This generalizes the well-known result that the chemical potential is the velocity of the phase for a stationary GP wavefunction, $\Psi(\mathbf{x},t)=\Psi_0(\mathbf{x})e^{-i\mu t}$, to an arbitrary number of spontaneously broken symmetries. 

Apart from a dynamical role, Lagrange multipliers also play a thermodynamical one, as they can be obtained from the usual energy $E=K+\lambda^\alpha Q_\alpha$, still conserved as $\{K,Q_\alpha\}$ are conserved charges. Evaluating its expression for $\Psi_0$ and making use of the GPG equation yields
\begin{eqnarray}\label{eq:EExpansion}
 \nonumber E&=&\int\mathrm{d}\mathbf{x}\,\left[-\frac{\Psi^*_0\nabla^2 \Psi_0}{2}+V(\mathbf{x})|\Psi_0|^2+\frac{|\Psi_0|^4}{2}\right]\\
&=& \int\mathrm{d}\mathbf{x}\left[\Psi_0^*H_{GP}\Psi_0-\frac{|\Psi_0|^4}{2}\right]\\
&=& \nonumber\lambda^\alpha Q_\alpha-\int\mathrm{d}\mathbf{x}\,\frac{|\Psi_0|^4}{2}.
\end{eqnarray}
In order to proceed further, we first take derivative with respect to the charge $Q_\alpha$ in the GPG equation,
\begin{equation}\label{eq:GPGGibbsMode} K_{GP}\partial_A\Psi_0+(\partial_A |\Psi_0|^2)\Psi_0=\partial_A\lambda^\beta T_\beta\Psi_0,
\end{equation}
and then evaluate the scalar product with $\Psi_0$, arriving at
\begin{equation}\label{eq:GPGDuhem}
\int\mathrm{d}\mathbf{x}\,\frac{\partial_A|\Psi_0|^4}{2}=Q_\beta \partial_A\lambda^\beta,
\end{equation}
where we have used that $K_{GP}$ is an Hermitian operator and $K_{GP}\Psi_0=0$. With the help of this result, we simply find
\begin{equation}\label{eq:GeneralChemicalPotential}
\partial_AE=\frac{\partial E}{\partial Q_\alpha}=\lambda^\alpha\equiv \lambda_A,
\end{equation}
where, for any vector $v$ with upper indices $\alpha$, we lower indices as $v_A\equiv v^\alpha$. This notation will be justified in Sec. \ref{subsec:BerryGibbs}, from where a simpler derivation of this result can be obtained [see Eq. (\ref{eq:ChargeDerivatives}) and ensuing discussion]:
\begin{align}
\nonumber \partial_A E&=\int\mathrm{d}\mathbf{x}\,\left[(\partial_A\Psi_0^*) H_{GP}\Psi_0+ (H_{GP}\Psi_0)^*\partial_A\Psi_0\right]\\
&=i\lambda^\beta(z_A|z_\beta)=\lambda^\alpha.
\end{align}

Interestingly, since $p=|\Psi_0|^4/2$ is the local pressure, we can rewrite Eq. (\ref{eq:EExpansion}) as
\begin{equation}
    E=\lambda^\alpha Q_\alpha - \int\mathrm{d}\mathbf{x}~p, 
\end{equation}
where we note that the conserved charges are macroscopic extensive magnitudes, resulting from integrals over the whole volume. By combining this expression with Eqs. (\ref{eq:GPGDuhem}), (\ref{eq:GeneralChemicalPotential}) and identifying the element of volume as $dV\equiv d\mathbf{x}$, we retrieve the generalization for non-homogeneous systems of the first principle of Thermodynamics (notice that we work at $T=0$),
\begin{equation}\label{eq:FirstThermo}
    dE=\lambda^\alpha d Q_\alpha-pdV,
\end{equation}
and of the Gibbs-Duhem relation, 
\begin{equation}\label{eq:GibbsDuhem}
    \int\mathrm{d}V~\partial_A p= Q_\beta \partial_A\lambda^\beta.
\end{equation}
Since the continuous symmetry parameters and their associated charges are conjugate position/momentum variables in a Hamiltonian language, these thermodynamic relations explain the role of the Lagrange multipliers as generalized velocities,
\begin{equation}\label{eq:HamiltonianFlow}
    \dot{\alpha}=\partial_A H=\lambda^\alpha.
\end{equation}
On the other hand, since the Hamiltonian is invariant under the symmetry transformations themselves, we immediately retrieve charge conservation, $\dot{Q}_\alpha=-\partial_\alpha H=0$.

\subsection{Nambu-Goldstone modes in the generalized Gibbs ensemble}\label{subsec:BdGG}
Once we have characterized the spontaneous symmetry-breaking states within the generalized Gibbs ensemble, we study the emergence of NG modes in their spectrum of excitations. For a condensate, these are described by the BdG equations. In a classical context, the BdG equations result from taking $\Psi'=\Psi+\varphi$ in the time-dependent GP equation (\ref{eq:GPEquation}) and expanding up to linear order in the field fluctuations $\varphi$,
\begin{equation}\label{eq:TimeDependentBdG} 
i\partial_t\Phi=M(t)\Phi,~~\Phi=\left[\begin{array}{c}\varphi\\ \varphi^*\end{array}\right],
\end{equation}
where
\begin{eqnarray}
\nonumber M(t)&=&\left[\begin{array}{cc} N(t) & A(t)\\
-A^*(t) &-N^*(t)\end{array}\right],\\
    N(t)&=&-\dfrac{\nabla^2}{2}+V(\mathbf{x})+2|\Psi(\mathbf{x},t)|^2,\\
    \nonumber A(t)&=&\Psi(\mathbf{x},t)^2.
\end{eqnarray}
In general, any continuous parameter $a$ on which the wavefunction depends, but the original Hamiltonian does not, has an associated time-dependent BdG solution:
\begin{equation}\label{eq:GeneralTimeDependentMode}
    i\partial_t z_a=M(t)z_a,~ z_a\equiv \left[\begin{array}{l}\partial_a\Psi\\ \partial_a\Psi^*\end{array}\right].
\end{equation}
This includes the spontaneously broken symmetries, 
\begin{equation}
    z_\alpha=\left[\begin{array}{l}\partial_\alpha\Psi\\ \partial_\alpha\Psi^*\end{array}\right]=\left[\begin{array}{l}-iT_\alpha\Psi\\ i(T_\alpha\Psi)^*\end{array}\right],
\end{equation}
since $\Psi_{\boldsymbol{\alpha}}=e^{-i\boldsymbol{\alpha}\cdot \mathbf{T}}\Psi$ is also a solution of the time-dependent GP equation.


Applying the same reasoning to $\Psi_0$ and the time-independent GPG equation (\ref{eq:GPGEquation}), we obtain that $z_\alpha$ is here a zero mode of the BdG-Gibbs (BdGG) equations:
\begin{equation}\label{eq:BdGGoldstone}
    M_0z_\alpha=0,~M_0=\left[\begin{array}{cc} N_0 & A_0\\
-A_0^* &-N_0^*\end{array}\right],
\end{equation}
where now
\begin{equation}\label{eq:BdGGMatrix}
    N_0=-\dfrac{\nabla^2}{2}+V(\mathbf{x})+2|\Psi_0(\mathbf{x})|^2-\boldsymbol{\lambda}\cdot \mathbf{T},~A_0=\Psi^2_0(\mathbf{x}).
\end{equation}
We can then identify $z_\alpha$ as the NG mode associated to the spontaneous symmetry breaking in a particular application of the general result of Eq. (\ref{eq:VariationalGoldstone}). In turn, each NG mode $z_\alpha$ is paired with a Gibbs mode $z_A$, obtained by taking derivative in the GPG equation with respect to the corresponding conserved charge $Q_\alpha$ [see Eq. (\ref{eq:GPGGibbsMode})],
\begin{equation}\label{eq:BdGGibbs}
    M_0z_A=i\partial_A\lambda^\beta z_\beta,
\end{equation}
which means that $z_A$ is a zero BdGG mode with a source term. Another way to put it is that the Gibbs modes are zero modes of the operator $M^2_0$, $M^2_0z_A=0$.

We note that the linear independence of the NG modes $z_\alpha$ is already guaranteed by the assumption of the linear independence between the components of the vector $\mathbf{T}\Psi_0$ [see discussion after Eq. (\ref{eq:GPGEquation})]. As an example, consider a one-dimensional (1D) plane wave solution $\Psi_0(x)=e^{iqx}$ in a homogeneous system, which apparently spontaneously breaks both $U(1)$ and spatial-translation symmetry. However, since $T_x\Psi_0=qT_\theta\Psi_0=q\Psi_0$, then $z_x= qz_\theta$, and we only have one independent NG mode. Indeed, the resulting BdG dispersion relation is that of a homogeneous system plus a Doppler shift \cite{Recati2009}. The momentum is also proportional to the particle number as $P=q N$. In general, when the system is in the ground state, there are counting rules relating the number of independent NG modes and broken-symmetry generators \cite{Watanabe2012,Hidaka2013,Watanabe2020}.


The BdGG dynamics is obtained by performing a unitary transformation on the GP wavefunction in Eq. (\ref{eq:GPEquation}),
\begin{equation}\label{eq:DynamicalGPExpansionClassical}
    \Psi(\mathbf{x},t)=e^{-i\boldsymbol{\lambda}\cdot \mathbf{T} t}\Psi'(\mathbf{x},t),
\end{equation}
which yields the time-dependent version of the GPG equation,
\begin{equation}\label{eq:TimeDependentGPG}
    i\partial_t\Psi'=K_{GP}\Psi'.
\end{equation}
Considering fluctuations around the stationary GPG wavefunction, $\Psi'(\mathbf{x},t)=\Psi_0(\mathbf{x})+\varphi(\mathbf{x},t)$, yields at linear order the time-dependent BdGG equations
\begin{equation}\label{eq:ClassicalBdGGFieldEquations}
i\partial_t \Phi=M_0 \Phi.
\end{equation}
The matrix $M_0$ is a time-independent linear operator, presenting eigenmodes of the form 
\begin{equation}
    M_0z_n=\epsilon_n z_n,~z_n\equiv \left[\begin{array}{c}u_n\\ v_n\end{array}\right]. %
\end{equation}
These modes are orthogonal under the inner product
\begin{equation}\label{eq:KGProduct}
    (z_n|z_m)\equiv\braket{z_n|\sigma_z z_m} =\int\mathrm{d}\mathbf{x}~[u^*_nu_m-v^*_nv_m], 
\end{equation}
since $M_0$ is pseudo-Hermitian, $(z_n|M_0z_m)=(M_0z_n|z_m)$, where $\braket{z_n|z_m}$ is the usual scalar product for two spinors and $\sigma_i$ are the Pauli matrices. The pseudo-Hermiticity of $M_0$ also implies that the inner product between two solutions of the time-dependent BdGG equation (\ref{eq:ClassicalBdGGFieldEquations}) is conserved. However, both $M_0$ and the inner product are not positive definite. Indeed, by defining conjugate modes as $\bar{z}_n\equiv \sigma_x z_n^*$, it is easy to see that 
\begin{eqnarray}
    M_0\bar{z}_n&=&-\epsilon^*_n\bar{z}_n,\\
    \nonumber (z_n|z_m)&=&-(\bar{z}_m|\bar{z}_n)=-(\bar{z}_n|\bar{z}_m)^*.
\end{eqnarray}
In the following, for simplicity, we assume that there are no dynamical instabilities (i.e., there are no complex eigenvalues) and that the only zero modes are the NG modes. It is immediate to show that one can separate the BdGG modes into two orthogonal sectors: the Goldstone-Gibbs sector formed by the paired modes $\{z_a\}=\{z_\alpha,z_A\}$, and the regular Bogoliubov sector formed by the eigenmodes $\{z_n\}$,
\begin{eqnarray}
    (z_n|z_m)&=&\delta_{nm},\\
    \nonumber (z_a|z_n)&=&0.
\end{eqnarray}

\subsection{Berry-Gibbs connection}\label{subsec:BerryGibbs}

The derivation of the orthogonality relations within the Goldstone-Gibbs sector is more delicate. First, we note that the modes $z_a$ are self-conjugates, $z_a=\bar{z}_a$, so their inner product is purely imaginary, $(z_a|z_b)=-(z_a|z_b)^*$, and they have zero norm, $(z_a|z_a)=0$. In particular, for the NG modes, since by assumption $[T_\alpha,T_\beta]=0$, $\braket{\Psi_0|[T_\alpha,T_\beta]|\Psi_0}=0$, and then 
\begin{equation}\label{eq:GoldstoneOrto}
(z_\alpha|z_\beta)=\braket{T_\alpha\Psi_0|T_\beta\Psi_0}-\braket{T_\beta\Psi_0|T_\alpha\Psi_0}=0
\end{equation}
as the generators $T_\alpha$ are Hermitian. In general, deriving a conserved charge with respect to any continuous parameter $b$ yields
\begin{eqnarray}\label{eq:ChargeDerivatives}
    \nonumber \partial_b Q_\alpha&=&\int\mathrm{d}\mathbf{x}~(T_\alpha\Psi_0)^*\partial_b\Psi_0+(\partial_b\Psi_0)^*T_\alpha\Psi_0\\
    &=&-i(z_\alpha|z_b).
\end{eqnarray}
Since the conserved charges are independent variables, we obtain that 
\begin{equation}\label{eq:GoldstoneGibbsOrto}
(z_\alpha|z_B)=i \partial_B Q_\alpha=i\delta_{AB}.    
\end{equation}

Finally, we only need to study the orthogonality between the Gibbs modes $z_A$. We show that one can indeed choose $(z_A|z_B)=0$ by performing a generalized gauge transformation as follows. The wavefunction $\Psi_{\boldsymbol{\chi}}=e^{-i\boldsymbol{\chi}\cdot \mathbf{T}}\Psi_0$ is also a solution of the GPG equation, where the vector $\boldsymbol{\chi}$ only depends on the conserved charges, $\boldsymbol{\chi}=\boldsymbol{\chi}(\mathbf{Q})$. Under the transformation $\Psi_0\to \Psi_{\boldsymbol{\chi}}$, 
the Gibbs modes $z_A$ and their inner product transform as
\begin{align}
    \partial_A\Psi_0&\to e^{-i\boldsymbol{\chi}\cdot \mathbf{T}}\left[\partial_A-i\partial_A\chi^\beta T_\beta \right]\Psi_0,\\
    \nonumber (z_A|z_B)&\to (z_A|z_B)+i(\partial_A\chi^\beta-\partial_B\chi^\alpha),
\end{align}
while the rest of the orthogonality relations remain unchanged. By defining a real antisymmetric tensor $F_{AB}\equiv i(z_A|z_B)$, we see from the equation above that we can set $(z_A|z_B)=0$ if we find a generalized gauge transformation satisfying
\begin{equation}
    F_{AB}=\partial_A\chi_B-\partial_B\chi_A,
\end{equation}
which means that $F$ is the exterior derivative of $\boldsymbol{\chi}$, $F=d\chi$. It is easy to check that $F_{AB}$ obeys the Bianchi identity
\begin{equation}
    \partial_C F_{AB}+\partial_B F_{CA}+\partial_A F_{BC}=0.
\end{equation}
Hence, the exterior derivative of $F$ vanishes, $dF=0$. This implies, via Poincar\'e's Lemma, that $F$ is exact, i.e., it can be written as the exterior derivative of some vector field $W$, $F=dW$. Thus, by directly taking $\chi_A=W_A$, we can always set $(z_A|z_B)=0$. In fact,
\begin{equation}\label{eq:BerryGibbsCurvature}
F_{AB}=i(\braket{\partial_A\Psi_0|\partial_B\Psi_0}-\braket{\partial_B\Psi_0|\partial_A\Psi_0})
\end{equation}
is nothing else but the usual Berry curvature evaluated for the GPG wavefunction, which here depends on the set of conserved charges $\mathbf{Q}$ (we recall that these are not intrinsic parameters of the original Hamiltonian). We then immediately identify $F=dW$, with $W_A=i\braket{\Psi_0|\partial_A\Psi_0}$\footnote{Strictly speaking, it should be $W_A=i(\braket{\Psi_0|\partial_A\Psi_0}-\braket{\partial_A\Psi_0|\Psi_0})/2$ to ensure that $W_A$ is real since $\partial_A \braket{\Psi_0|\Psi_0}=\partial_A N\neq 0$.}. As a result of this analogy, we denote the vector $W$ as the Berry-Gibbs connection and $F_{AB}$ as the Berry-Gibbs curvature. Its main differences with the standard Berry connection are i) $\braket{\Psi_0|\Psi_0}$ does depend on the connection parameters (specifically, $\braket{\Psi_0|\Psi_0}=N$); and ii) we can perform generalized gauge transformations beyond the usual phase transformations $\Psi_0\to e^{-i\theta(\mathbf{Q})}\Psi_0$, under which $F_{AB}$ is not invariant.

The Berry-Gibbs curvature can be extended to the complete Goldstone-Gibbs manifold with coordinates $x^a=(x^\alpha,x^A)=(\boldsymbol{\alpha},\mathbf{Q})$ as 
\begin{equation}\label{eq:BerryGibbsCurvatureExtended}
    F_{ab}\equiv i(z_a|z_b)=i(\braket{\partial_a\Psi_0|\partial_b\Psi_0}-\braket{\partial_b\Psi_0|\partial_a\Psi_0}),
\end{equation}
where we recall that $\partial_\alpha\Psi_0=-iT_\alpha\Psi_0$. In matrix notation, it reads
\begin{equation}
    F=\left[\begin{array}{cc} F_{\alpha\beta} & -I_n\\
I_n & F_{AB}\end{array}\right], 
\end{equation}
$I_n$ being the $n\times n$ identity matrix. The off-diagonal blocks are determined by Eq. (\ref{eq:GoldstoneGibbsOrto}). In this work, we have assumed that the generators of the broken symmetries commute between themselves. This implies that $F_{\alpha\beta}=0$, Eq. (\ref{eq:GoldstoneOrto}), and that one can always set $F_{AB}=0$ by means of a generalized gauge transformation, as discussed above. Actually, within the complete Goldstone-Gibbs manifold, a generalized gauge transformation is just a change of coordinates 
\begin{equation}  x'^\alpha=x^\alpha+\chi^\alpha(\mathbf{Q}),~x'^A=x^A=Q_\alpha.
\end{equation}

We can then summarize all the orthogonality relations as
\begin{eqnarray}\label{eq:OrthogonalityBdGG}
    \nonumber (z_n|z_m)&=&\delta_{nm},\\
    (z_a|z_n)&=&0,\\
    \nonumber (z_a|z_b)&=&-iF_{ab}=i\Omega_{ab},
\end{eqnarray}
with $\Omega$ the $2n\times 2n$ symplectic form,
\begin{equation}
    \Omega=\left[\begin{array}{cc} 0 & I_n\\
-I_n & 0\end{array}\right].
\end{equation}
This symplectic form allows us to lower (rise) indices as
\begin{equation}
    x_b=x^a\Omega_{ab},~x^a=x_b\Omega^{ba},
\end{equation}
where $\Omega^{ab}$ is the inverse of the symplectic matrix, $\Omega^{-1}=-\Omega=\Omega^T$, satisfying $\Omega_{ac}\Omega^{cb}=\Omega^{bc}\Omega_{ca}=\delta^b_a
$. 





\subsection{Quantization}\label{subsec:Quantum}

When returning to the quantum theory, quantization \`a la Dirac prescribes that Poisson brackets are promoted to commutators as $[\hat{A},\hat{B}]=i\hbar \{A,B\}$. Therefore, the field and the conserved charges are promoted to quantum operators which satisfy [see Eqs. (\ref{eq:CanonicalCommutationClassical}), (\ref{eq:ChargeGenerationClassical})]
\begin{align}\label{eq:CanonicalCommutationQuantum}
    [\hat{\Psi}(\mathbf{x}),\hat{\Psi}^\dagger(\mathbf{x}')]&=\delta(\mathbf{x}-\mathbf{x}'),\\
   \nonumber [\hat{\Psi}(\mathbf{x}),\hat{Q}_\alpha]&=T_\alpha \hat{\Psi}(\mathbf{x}).
\end{align}
In the case of a fermionic field operator, the commutator in the first line is replaced by an anticommutator. Symmetry transformations on the field operator can be implemented as
\begin{align}
     e^{i \boldsymbol{\alpha}\cdot \mathbf{\hat{Q}}}\hat{\Psi}(\mathbf{x}) e^{-i \boldsymbol{\alpha}\cdot \mathbf{\hat{Q}}}&=e^{-i\boldsymbol{\alpha}\cdot \mathbf{T}}\hat{\Psi}(\mathbf{x}),
\end{align}
where $\mathbf{\hat{Q}}$ is a vector containing the second-quantized charges $\hat{Q}_\alpha$, in turn satisfying $[\hat{H},\hat{Q}_\alpha]=[\hat{Q}_\alpha,\hat{Q}_\beta]=0$. The  equation of motion for the quantum field operator is precisely the Heisenberg equation (\ref{eq:HeisenbergEquationOfMotion}).

The Lagrangian description of Eq. (\ref{eq:LagrangianGibbs}) is retrieved by performing a unitary transformation
\begin{equation}\label{eq:GibbsPicture}
    \hat{\Psi}(\mathbf{x},t)\equiv e^{i\boldsymbol{\lambda}\cdot\mathbf{\hat{Q}}t}\hat{\Psi}'(\mathbf{x},t)e^{-i\boldsymbol{\lambda}\cdot\mathbf{\hat{Q}}t}= e^{-i\boldsymbol{\lambda}\cdot \mathbf{T} t} \hat{\Psi}'(\mathbf{x},t),
\end{equation}
equivalent to working in a Gibbs picture where the dynamics is governed by the generalized Gibbs Hamiltonian, $\hat{\Psi}'(\mathbf{x},t)=e^{i\hat{K}t}\hat{\Psi}(\mathbf{x})e^{-i\hat{K}t}$. The point of the Gibbs picture is that it amounts to switching to a rotating frame where the symmetry-breaking state is time-independent, and the dynamics of its quantum fluctuations is determined by a stationary problem.

\begin{widetext}

These results are so far general and exact, easily translatable to an arbitrary many-body system. In the specific case of a condensate, the Heisenberg equation of motion is the quantum version of the time-dependent GP equation (\ref{eq:GPEquation}),
\begin{equation}\label{eq:HeisenbergEquationOfMotionBoson}
    i\partial_t\hat{\Psi}(\mathbf{x},t)=\left[-\frac{\nabla^2}{2}+V(\mathbf{x})+\hat{\Psi}^{\dagger}(\mathbf{x},t)\hat{\Psi}(\mathbf{x},t)\right]\hat{\Psi}(\mathbf{x},t),
\end{equation}
while the Gibbs picture yields the quantum version of the time-dependent GPG equation (\ref{eq:TimeDependentGPG}),
\begin{equation}\label{eq:HeisenbergGibbsEquationOfMotion}
i\hbar\partial_t\hat{\Psi}'(\mathbf{x},t)=[\hat{\Psi}'(\mathbf{x},t),\hat{K}]=\left[-\frac{\nabla^2}{2}+V(\mathbf{x})+\hat{\Psi}'^{\dagger}(\mathbf{x},t)\hat{\Psi}'(\mathbf{x},t)-\boldsymbol{\lambda}\cdot\mathbf{T}\right]\hat{\Psi}'(\mathbf{x},t).
\end{equation} 
\end{widetext}

Quantum fluctuations around the coherent GPG expectation value can be described by expanding the field operator as
$\hat{\Psi}'(\mathbf{x},t)=\Psi_0(\mathbf{x})+\hat{\varphi}(\mathbf{x},t)$, recovering at linear order the quantum time-dependent BdGG equations (\ref{eq:ClassicalBdGGFieldEquations}),
\begin{equation}\label{eq:QuantumBdGGFieldEquations}
i\partial_t \hat{\Phi}=M_0 \hat{\Phi},~\hat{\Phi}(\mathbf{x},t)=\left[\begin{array}{c}\hat{\varphi}(\mathbf{x},t)\\ \hat{\varphi}^{\dagger}(\mathbf{x},t)\end{array}\right].
\end{equation}
Using that $\{z_a,z_n\}$ form a complete of set of modes, we can expand the quantum fluctuations of the field operator as
\begin{equation}\label{eq:QuantumFieldFluctuations}
   \hat{\Phi}(\mathbf{x},t)=\hat{\gamma}^{a}(t)z_a(\mathbf{x})+\sum_{n}\hat{\gamma}_{n}(t)z_{n}(\mathbf{x})+\hat{\gamma}^{\dagger}_{n}(t)\bar{z}_{n}(\mathbf{x}),
\end{equation}
where $\hat{\gamma}^a,\hat{\gamma}_n$ are the quantum amplitudes of each BdGG mode. In the case of regular Bogoliubov modes, their amplitude is obtained from
\begin{equation}
    \hat{\gamma}_n(t)\equiv (z_n|\hat{\Phi}(t)).
\end{equation}
By invoking the canonical commutation rules for the field operator, first line in Eq. (\ref{eq:CanonicalCommutationQuantum}), it is shown that these amplitudes behave as bosonic annhihilation operators,
\begin{equation}
    [\hat{\gamma}_n,\hat{\gamma}^\dagger_m]=[(z_n|\hat{\Phi}),(\hat{\Phi}|z_m)]=(z_n|z_m)=\delta_{nm}.
\end{equation}
The corresponding equation of motion is simply
\begin{eqnarray}\label{eq:BogoliubovQuantum}
    \nonumber i\partial_t \hat{\gamma}_n(t)&=&(z_n|M_0\hat{\Phi})=\epsilon_n\hat{\gamma}_n(t)\Longrightarrow \\\hat{\gamma}_n(t)&=&\hat{\gamma}_ne^{-i\epsilon_nt}.
\end{eqnarray}

Regarding the amplitudes of the Goldstone-Gibbs modes, quantization is more tricky because of their zero norm. In this case, we have that
\begin{equation}
(z_a|\hat{\Phi})=i\Omega_{ab}\hat{\gamma}^b=- i \hat{\gamma}_a\Longrightarrow \hat{\gamma}_a=i(z_a|\hat{\Phi}).
\end{equation}
Since both $z_a,\hat{\Phi}$ are self-conjugates, $\hat{\gamma}_a$ is a Hermitian operator, $\hat{\gamma}_a=\hat{\gamma}^\dagger_a$, and so is $\hat{\gamma}^a=\hat{\gamma}_b\Omega^{ba}$. Their commutation relations are readily found to be
\begin{equation}
    [\hat{\gamma}_a,\hat{\gamma}_b]=(z_a|z_b)=i\Omega_{ab} \Longrightarrow [\hat{\gamma}^a,\hat{\gamma}^b]=-i\Omega^{ab}.
\end{equation}

This implies that the components of the vector $\boldsymbol{\hat{\gamma}}=(\hat{\gamma}^\alpha,\hat{\gamma}^A)\equiv (\hat{X}^\alpha,\hat{P}^A)$ can be seen as generalized coordinate/momentum operators, as those also satisfy the same symplectic algebra:
\begin{equation}\label{eq:CoordinateMomentum}
    [\hat{X}^\alpha,\hat{P}^B]=[\hat{X}_A,\hat{P}^B]=i\delta_A^{B}.
\end{equation}
Indeed, we can take this analogy further by analyzing their equations of motion,
\begin{equation}
    i\partial_t \hat{\gamma}_a(t)=i(z_a|M_0\hat{\Phi})=i(M_0z_a|\hat{\Phi}).
\end{equation}
Specifically,
\begin{align} \hat{X}^\alpha&=\hat{\gamma}^\alpha=i(z_A|\hat{\Phi}),\\
\nonumber \hat{P}^A&=\hat{\gamma}^A=-i(z_\alpha|\hat{\Phi}),   
\end{align}
and thus, from Eqs. (\ref{eq:BdGGoldstone}), (\ref{eq:BdGGibbs}), we get
\begin{align}\label{eq:GoldstoneGibbsQuantum}
    i\partial_t\hat{P}^A&=0\Longrightarrow \hat{P}^A(t)=\hat{P}^A,\\ \nonumber i\partial_t\hat{X}^\alpha&=\partial_A\lambda^\beta(z_\beta|\hat{\Phi})=i \partial_A\lambda_B \hat{P}^B\Longrightarrow \\ \nonumber \hat{X}^\alpha(t)&=\hat{X}_A(t)=\hat{X}_A+(\partial_A\lambda_B) \hat{P}^B t.
\end{align}
In this way, the conserved momenta $\hat{P}^B$ yield the velocities of the coordinates $\hat{X}_A$. Moreover, since in turn $\lambda_A=\partial_A E$, then $\partial_A\lambda_B=\partial_B\lambda_A=\partial^2_{AB}E$, and we obtain the meaningful relation
\begin{equation}\label{eq:QuantumGoldstoneFluctuations}
    \hat{X}_A(t)=\hat{X}_A+(\partial_A\lambda_B) \hat{P}^B t=\hat{X}_A+(\partial_B\lambda_A) \hat{P}^B t.
\end{equation}
The physics behind this equation is very simple. After quantization, each conserved charge is expanded up to linear order in the field fluctuations in the same fashion as Eq. (\ref{eq:ChargeDerivatives}),
\begin{equation}
    \hat{Q}_\alpha\simeq Q_\alpha-i(z_\alpha|\hat{\Phi})=Q_\alpha+\hat{P}^A, 
\end{equation}
where $Q_\alpha$ is the mean-field value of the charge. Thus, the momentum $\hat{P}^A$ describes the quantum fluctuations of the associated charge, $\delta  \hat{Q}_\alpha\simeq \hat{P}^A $, from where it inherits its time-independent character. These quantum fluctuations precisely stem from the fact that the spontaneous breaking of a given symmetry implies that the system cannot be in an eigenstate of the associated charge. On the other hand, the GPG wavefunction displaces with constant velocity along the orbits generated by the broken-symmetry transformations, Eq. (\ref{eq:TimeDependentGPGEquation}), whose tangent space is precisely spanned by the NG modes. In particular, the velocity component along $z_\alpha$ is the Lagrange multiplier $\lambda_A$, which depends on the conserved charges $\lambda_A=\lambda_A(\mathbf{Q})$. In the quantum theory, these velocities also fluctuate around their mean-field values as the conserved charges do: 
\begin{equation}\label{eq:LagrangeQF}
    \hat{\lambda}_A=\lambda_A(\mathbf{\hat{Q}})\simeq \lambda_A(\mathbf{Q})+\partial_B\lambda_A \hat{P}^B.
\end{equation}
The quantum fluctuations of the velocities are then translated into a ballistic motion for the quantum amplitude of the NG modes, arriving at Eq. (\ref{eq:QuantumGoldstoneFluctuations}). Another way to put it is that the modes $z_A$ are not eigenmodes of the BdGG operator $M_0$. Instead, from Eq. (\ref{eq:TimeDependentGPGEquation}), we have that
\begin{equation} 
\partial_A\Psi(\mathbf{x},t)=e^{-i\boldsymbol{\lambda}\cdot \mathbf{T} t}\left[\partial_A-i(\partial_A\lambda^\beta)T_\beta\,t\right]\Psi_0,
\end{equation}
which yields $z_A(\mathbf{x},t)=z_{A}(\mathbf{x})+(\partial_A\lambda^\beta)z_\beta t$ in the time-dependent BdGG equation (\ref{eq:ClassicalBdGGFieldEquations}). As a result, we recover the same intuitive picture of the classical mechanical example in Sec. \ref{sec:GoldstoneGeneral}: the dynamics of the fluctuations along the orbits generated by the broken symmetry transformations lead to unbounded ballistic motion as there is no restoring force. These results are thus limited to sufficiently short times, when the NG amplitudes $\hat{X}^\alpha(t)$ can be treated as small, and the linear BdG approximation is still valid. Nevertheless, our formalism also provides the basis for a full nonperturbative expansion \cite{Lewenstein1996,Dziarmaga2004}, whose study is beyond the scope of the present work.


After gathering Eqs. (\ref{eq:BogoliubovQuantum}), (\ref{eq:GoldstoneGibbsQuantum}), we find that the quantum field fluctuations evolve as
\begin{align}\label{eq:QuantumFieldFluctuationsExplicit}
   \nonumber &\hat{\Phi}(\mathbf{x},t)=[\hat{X}^\alpha+\partial_A\lambda_B \hat{P}^B t] z_\alpha(\mathbf{x})+ \hat{P}^Az_A(\mathbf{x})\\
   &+\sum_{n}\hat{\gamma}_{n}z_{n}(\mathbf{x})e^{-i\epsilon_nt}+\hat{\gamma}^{\dagger}_{n}(t)\bar{z}_{n}(\mathbf{x})e^{i\epsilon_nt}.
\end{align}
The same dynamics could have been derived from the generalized Gibbs Hamiltonian by inserting Eq. (\ref{eq:QuantumFieldFluctuations}) in $\hat{K}=\hat{H}-\boldsymbol{\lambda}\cdot \mathbf{\hat{Q}}$ and expanding up to quadratic order in the field fluctuations, in the spirit of the Bogoliubov approximation. After dropping out all mean-field and zero-point $c$-number contributions, we arrive at
\begin{equation}
      \hat{K}=\frac{1}{2}(\hat{\Phi}|M_0\hat{\Phi})=\frac{\partial^2_{AB}E}{2}\hat{P}^A\hat{P}^B+\sum_n \epsilon_n \hat{\gamma}^{\dagger}_{n}\hat{\gamma}_{n}.
\end{equation}
The second term is the usual Bogoliubov contribution while the first one is the Goldstone-Gibbs contribution. The latter is quadratic in the fluctuations of the conserved charges, then behaving as effective momenta, and does not depend on the NG amplitudes, since the Hamiltonian is invariant under the symmetry transformations. In particular, if the solution is energetically stable, then the susceptibility matrix $\partial^2_{AB}E$ is a positive definite form. In fact, one can always diagonalize $\partial^2_{AB}E$ by performing an appropriate rotation in the Goldstone-Gibbs sector, $\partial^2_{AB}E=M_A^{-1}\delta_{AB}$, where $M_A$ plays the role of an effective mass. This rotation yields a new set of independent Goldstone-Gibbs modes $\{z'_\alpha,z'_A\}$, whose amplitudes $\{\hat{X}'^\alpha,\hat{P}'^A\}$ satisfy   
\begin{equation}\label{eq:HybridNGModes}
      \hat{K}=\frac{(\hat{P}'^A)^2}{2M_A}+\sum_n \epsilon_n \hat{\gamma}^{\dagger}_{n}\hat{\gamma}_{n},~\hat{X}'^\alpha(t)=\hat{X}'^\alpha+\frac{\hat{P}'^A}{M_A}t.
\end{equation}
Thus, the susceptibility matrix, computed from thermodynamical considerations, determines how the different NG modes hybridize in a dynamical context. Moreover, as the NG modes are the zero-energy limit of some branches from the regular Bogoliubov sector, the structure of the hybrid Goldstone-Gibbs modes $\{z'_\alpha,z'_A\}$ also describes the low-energy regime.

The quantization of the Goldstone-Gibbs modes in the usual GP context is simply retrieved by first setting all Lagrange multipliers to zero except for the chemical potential $\lambda^\theta=\mu$, so the only independent conserved charge is the particle number $N$. With the resulting GP solution, one computes the remaining charges $Q_\alpha$ to assemble the vector $\mathbf{Q}=\mathbf{Q}_0$. Once known, the full GPG equation is solved as a function of $\mathbf{Q}$, from where the Berry-Gibbs curvature $F_{AB}$ and the susceptibility matrix $\partial^2_{AB}E$ can be locally evaluated at the point $\mathbf{Q}_0$ characterizing the original GP solution. 

The presence of dynamical instabilities would not modify the main results of this section as these modes are orthogonal to both the Goldstone-Gibbs and the regular Bogoliubov sector. Their quantization is more involved as they behave like unstable oscillators, bearing some resemblance to the Goldstone-Gibbs procedure; see Refs. \cite{Leonhardt2003,Finazzi2010,deNova2024} for more details.

To conclude this section, we stress that the thermodynamical description is essential to ensure the conservation of the commutation rules between the NG amplitudes because it implies that the flow of the GPG wavefunction is potential, Eq. (\ref{eq:HamiltonianFlow}), and hence
\begin{equation}\label{eq:CommutationPreservation}
    [\hat{X}_A(t),\hat{X}_B(t)]=i(\partial_B \lambda_A-\partial_A \lambda_B)t=0. 
\end{equation}

\begin{figure}[tb!]
    \includegraphics[width=\columnwidth]{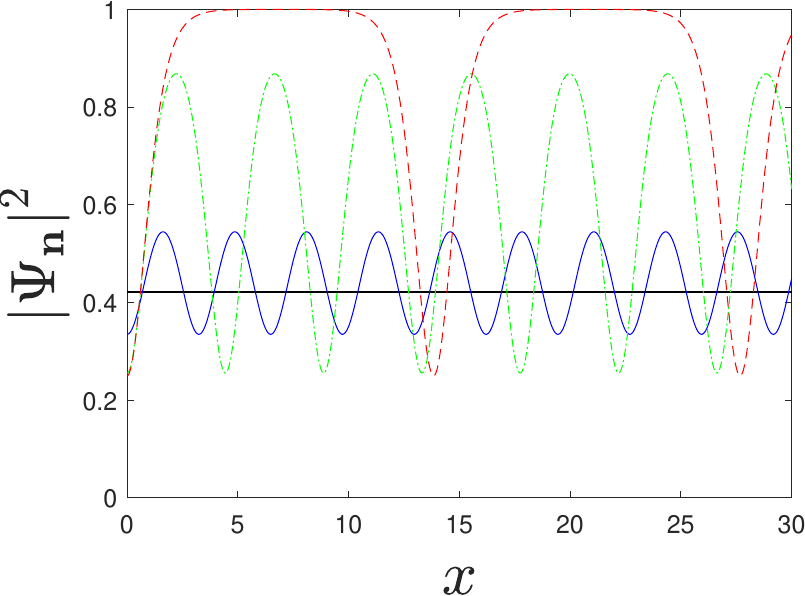} 
\caption{Density profile $|\Psi_{\mathbf{n}}(x)|^2$ of a cnoidal wave for increasing $\nu=0$ (horizontal black), $\nu=0.2$ (solid blue), $\nu=0.7$ (dashed-dotted green), and $\nu=0.9999$ (dashed red), with fixed $J=0.5$ and $\mu_v=1.125$.} 
\label{fig:Cnoidal}
\end{figure}

\section{Cnoidal wave in a superfluid}\label{sec:Cnoidal}


As a particular application of the formalism developed in the previous section, we consider a 1D quasi-condensate \cite{Menotti2002} in a ring of length $L$ with no external potential, $V(x)=0$, so there is invariance under spatial translations. This type of model has been recently used to understand certain aspects of supersolidity due to its appealing simplicity \cite{Martone2021}. We now look for solutions which simultaneously break both $U(1)$ and space-translation symmetry. In our notation, this amounts to taking $\boldsymbol{\alpha}=(\theta,x)$, $\mathbf{T}=(1,-i\partial_x)$ and $\mathbf{Q}=(N,P)$, respectively. The Lagrange multipliers associated to the particle number $N$ and momentum $P$ are $\lambda^\theta=\mu$ and $\lambda^x=v$, i.e., $\boldsymbol{\lambda}=(\mu,v)$. The resulting 
GPG equation (\ref{eq:GPGEquation}) then reads  
\begin{equation}\label{eq:GPGSupersolid}
    H_{GP}\Psi_0=\left[-\frac{\partial_x^2}{2}+|\Psi_0|^2\right]\Psi_0=\mu\Psi_0-iv\partial_x\Psi_0.
\end{equation}
We can get rid of $v$ by the transformation $\Psi_0(x)=e^{ivx}\tilde{\Psi}_0(x)$, which yields the standard homogeneous GP equation with a chemical potential $\mu_v$,
\begin{equation}\label{eq:GPGalileo}
H_{GP}\tilde{\Psi}_0=\left(\mu+\frac{v^2}{2}\right)\tilde{\Psi}_0\equiv \mu_v \tilde{\Psi}_0.
\end{equation}
The solutions to this equation are the celebrated cnoidal waves, 
$\tilde{\Psi}_0(x)=\Psi_{\mathbf{n}}(x)$, which are determined by a vector $\mathbf{n}=[n_1,n_2,n_3]$, $0\leq n_1\leq n_2\leq n_3$:
\begin{widetext}
    \begin{eqnarray}\label{eq:CnoidalWave}
\Psi_{\mathbf{n}}(x)&=&\sqrt{n(x)}e^{i\theta(x)},\\
\nonumber n(x)&=&n_1+(n_2-n_1)\text{sn}^2(\sqrt{n_3-n_1}x,\nu),\\
\nonumber \theta(x)&=&\int^x_0\frac{J}{n(x')}\mathrm{d}x'=\frac{J}{ n_1\sqrt{n_3-n_1}}\Pi\left(\text{am}(\sqrt{n_3-n_1}x,\nu),m,\nu\right),
\end{eqnarray}
\end{widetext}

where the elliptic parameters $ 0\leq\nu\leq 1$, $m\leq 0$ are
\begin{equation}\label{eq:EllipticParameters}
   \nu=\frac{n_2-n_1}{n_3-n_1},~m=1-\frac{n_2}{n_1},
\end{equation}
while the chemical potential $\mu_v$ and cnoidal current $J$ read
\begin{equation}
    \mu_v=\frac{n_1+n_2+n_3}{2},~J=\sqrt{n_1n_2n_3}.
\end{equation}
Technical details behind the calculations of this section as well as basic properties of elliptic functions are presented in the Appendix. 


Cnoidal waves are periodic solutions of a nonlinear wave equation, generalizing the usual linear sinusoidal waves. The simplest example is the oscillation of a pendulum, which can be analytically obtained in terms of elliptic functions [see Eq. (\ref{eq:PendulumAnalytical}) and ensuing discussion]. In our case, for fixed $J,\mu_v$, the parameter $\nu$ controls both the periodicity and the nonlinearity of the wave; see Fig. \ref{fig:Cnoidal}. In the linear limit $\nu\to 0$, we recover a small amplitude sinusoidal (horizontal black and solid blue lines), $\textrm{sn}(u,0)=\sin u$ and  $K(0)=\pi/2$, as in the case of the small oscillations of a pendulum around its stable equilibrium position. For increasing $\nu$, the oscillation amplitude and  periodicity grow (dashed-dotted green and dashed red); in the limit $\nu\to 1$, the period diverges as $\textrm{sn}(u,1)=\tanh u$ and $K(1)=\infty$. This is akin to the falling trajectory of a pendulum whose starting point is close to the unstable equilibrium position.

Specifically, the period $a$ of the cnoidal density $n(x)$ is
\begin{equation}
    a=\frac{2K(\nu)}{\sqrt{n_3-n_1}}.
\end{equation}
Regarding the phase $\theta(x)$, since $n(x)$ is periodic, $J/n(x)$ is periodic. The integral of a periodic function is a periodic function modulo a constant slope $w_{\mathbf{n}}x$, given by its zero Fourier component. By taking $\theta(0)=0$, we have
\begin{align}
    w_{\mathbf{n}}&=\frac{\theta(a)}{a}=\frac{J}{n_1}\frac{\Pi(m,\nu)}{K(\nu)}.
\end{align}
Hence, we separate the phase as 
\begin{equation}\label{eq:PeriodicCnoidalPhase}
\theta(x)=w_{\mathbf{n}}x+\int^x_0\left[\frac{J}{n(x')}-w_{\mathbf{n}}\right]\mathrm{d}x'\equiv w_{\mathbf{n}}x+\Theta(x),
\end{equation}
with $\Theta(x+a)=\Theta(x)$ and $\Theta(0)=0$. As a result, the total GPG wavefunction takes the form of a Bloch wave,
\begin{equation}\label{eq:BlochWaveCnoidal}
    \Psi_0(x)=e^{ivx}\Psi_{\mathbf{n}}(x)\equiv e^{i(v+w_{\mathbf{n}})x}u_{\mathbf{n}}(x),
\end{equation}
where $u_{\mathbf{n}}(x)$ is a periodic function, 
$u_{\mathbf{n}}(x+a)=u_{\mathbf{n}}(x)$, and $v+w_{\mathbf{n}}$ is the crystal momentum modulo $2\pi/a$.

We now impose the physical constraints that determine the $5$ independent parameters of the problem: $\mathbf{n},v,\mu$. The periodic boundary conditions of the ring $\Psi_0(x+L)=\Psi_0(x)$ yield two equations, one for the density and one for the phase. When joined by the particle number and momentum conservation as well as the relation between $\mathbf{n}$ and $\mu_v$, we arrive at a system of $5$ coupled equations:
\begin{eqnarray}\label{eq:CnoidalMatching5}
    L&=&\ell a=\frac{2  K(\nu) \ell}{\sqrt{n_3-n_1}},~\ell\in\mathbb{N},\\
    \nonumber \frac{2\pi q}{L} &=&v+w_{\mathbf{n}},~q\in\mathbb{Z},\\ 
    \nonumber N&=&\int^L_0\mathrm{d}x~|\Psi_0(x)|^2,\\
    \nonumber P&=&\int^L_0\mathrm{d}x~\Psi^*_0(-i\partial_x)\Psi_0,\\
    \nonumber \mu_v&=&\frac{n_1+n_2+n_3}{2},
\end{eqnarray} 
with $\ell\in\mathbb{N}$ the number of cnoidal periods inside the ring and $q\in\mathbb{Z}$ the winding number. For given $N,P,L$, only a finite number of values $(\ell,q)$ is possible, each one characterizing a different GPG solution. By noticing that the equation for $\mu_v$ is trivial and combining the momentum conservation plus the equation for the winding number, a closed system of $3$ equations is obtained for the vector $\mathbf{n}$ [see Eq. (\ref{eq:CnoidalMomentum}) and related calculations], 
\begin{eqnarray}\label{eq:ClosedCnoidalSystem}
    L&=&\frac{2 K(\nu) \ell}{\sqrt{n_3-n_1}},\\
     \nonumber \bar{n}&=&n_1+(n_3-n_1)\left[1-\frac{E(\nu)}{K(\nu)}\right],\\
     \nonumber \frac{2\pi q}{L}&=&\frac{\bar{p}}{\bar{n}}+\sqrt{n_1n_2n_3}\left[\frac{\Pi(m,\nu)}{n_1K(\nu)}-\frac{1}{\bar{n}} \right].
\end{eqnarray}
Remarkably, these expressions only involve intensive thermodynamic magnitudes:
\begin{equation}
    \bar{n}\equiv \frac{N}{L},~\bar{p}\equiv \dfrac{P}{L}.
\end{equation}
The procedure to numerically solve these equations is explained after Eq. (\ref{eq:CnoidalMomentum}). Once done, we evaluate the energy density by
\begin{eqnarray}
    e&\equiv &\frac{E}{L}=\mu \bar{n}+v\bar{p}+\frac{n_1^2}{2}-n_1\bar{n}\\
    \nonumber &-&\frac{(n_3-n_1)^2}{6}\left[(\nu+2)-2(\nu+1)\frac{E(\nu)}{K(\nu)}\right],
\end{eqnarray}
and numerically compute the derivatives $\partial_N E=\partial_{\bar{n}}e$, $\partial_P E=\partial_{\bar{p}}e$, checking that indeed we recover $\partial_N E=\mu,~\partial_P E=v$. We note that the above formalism can be straightforwardly adapted to account for the presence of a flux within the ring.


Due to the symmetry of the problem, if $\Psi_0(x)$ is a solution of the GPG equation, then $e^{-i\theta}\Psi_0(x-x_0)$ is also a solution of the GPG equation. This results in the Goldstone-Gibbs modes
\begin{align}
    \nonumber z_\theta(x)&=\left[\begin{array}{r}-i\Psi_0(x)\\ i\Psi_0^*(x)\end{array}\right],~z_x(x)=-\left[\begin{array}{l}\partial_x\Psi_0(x)\\ \partial_x\Psi_0^*(x)\end{array}\right],\\
    \nonumber z_N(x)&=\left[\begin{array}{l}\partial_N\Psi_0(x)\\ \partial_N\Psi_0^*(x)\end{array}\right],\,z_P(x)=\left[\begin{array}{l}\partial_P\Psi_0(x)\\ \partial_P\Psi_0^*(x)\end{array}\right].\\
\end{align}
We have explicitly evaluated the extended Berry-Gibbs curvature $F_{ab}=i(z_a|z_b)$, recovering the predicted symplectic form, $F_{ab}=-\Omega_{ab}$. Specifically, in our gauge choice, the Berry-Gibbs curvature automatically vanishes without the need of further generalized gauge transformations, $F_{NP}=F_{PN}=0$ [see Eq. (\ref{eq:BerryGibbsCnoidal}) and ensuing discussion]. In this example, a generalized gauge transformation amounts to adding global phase and spatial shifts that depend on the particle number and momentum, $\theta=\theta(N,P),~x_0=x_0(N,P)$. The quantum amplitudes of the NG modes can be then understood as the quantum fluctuations of the global phase and position of the cnoidal wave, $\hat{X}^\theta=\delta\hat{\theta}$, $\hat{X}^x=\delta\hat{x}_0$, while their momenta describe total particle number and momentum fluctuations, $\hat{P}^N=\delta\hat{N}$, $\hat{P}^P=\delta\hat{P}$. 

Regarding thermodynamics, in all numerical cases we have found that the susceptibility matrix $\partial^2_{AB}E$ is both non-diagonal and non-positive definite. This reflects, respectively, that phase and crystalline modes hybridize in the low-energy limit, and the presence of a negative-energy branch in the regular Bogoliubov sector as the true energy minimum is always a homogeneous GP solution \cite{Martone2021}. In fact, the results of Ref. \cite{Martone2021}, derived within the standard mean-field GP approach, correspond to setting $v=0$ instead of fixing the total momentum $P$, and taking the limit $L\to \infty$. However, even for $v=0$, the full GPG equation is still needed for the computation of the susceptibility matrix and the subsequent characterization of the Goldstone-Gibbs modes [see discussion after Eq. (\ref{eq:HybridNGModes})].


When inserting the stationary GPG wavefunction into the time-dependent GP equation, $\Psi(x,0)=\Psi_0(x)$, we obtain a traveling cnoidal wave,
\begin{equation}
    i\partial_t\Psi=H_{GP}\Psi \Longrightarrow \Psi(x,t)=\Psi_0(x-vt)e^{-i\mu t},
\end{equation}
which is equivalent to perform a Galilean transformation with velocity $v$ to the cnoidal wave solution of the time-independent homogeneous GP equation with chemical potential $\mu_v$. This explicitly shows that the Lagrange multiplier associated to the momentum $P$ is the velocity of the spatial translation of the condensate, $\dot{x}_0=v$, motivating the choice of the label $v$.

Naively, one could claim that the GPG wavefunction also behaves dynamically as a continuous time crystal since it is time periodic,
\begin{equation}\label{eq:AccidentalSFstate}
\Psi(x,t)=e^{i(v+w_{\mathbf{n}})x}u_{\mathbf{n}}(x-vt)e^{-i\tilde{\mu}t},
\end{equation}
with a period $T=a/v$ and a quasi-chemical potential $\tilde{\mu}=\mu+vw_{\mathbf{n}}+v^2$, defined modulo $\omega=2\pi/T$. However, this time-periodic behavior does not stem from a genuine spontaneous symmetry breaking of time-translation invariance but rather from that of space-translation invariance plus a constant motion along a closed broken-symmetry orbit, as set by a non-vanishing $v$, in the same fashion of Eq. (\ref{eq:TimeDependentGPGEquation}). Indeed, the NG mode arising from the alleged time-translation symmetry breaking is not independent from those of phase and spatial translations since $\partial_t\Psi=-i\mu\Psi-v\partial_x\Psi$, which implies $z_t=\mu z_\theta+v z_x$. This is also reflected by the energy dependence on the other conserved charges, $E=E(N,P)$. In general, for stationary GPG solutions, time-translation symmetry is not spontaneously broken by itself but rather in terms of other symmetries as $\partial_t\Psi=-i(\boldsymbol{\lambda}\cdot \mathbf{T})\Psi$ and $E=E(\mathbf{Q})$.

This analysis suggests that certain time-periodic systems that do not genuinely break time-translation symmetry can be misidentified as continuous time crystals, with the time periodicity simply arising from constant motion along a closed orbit spawned by other broken-symmetry generators.

\section{Variational Floquet-Goldstone Theorem: Floquet-Nambu-Goldstone modes}\label{sec:GoldstoneFloquetGeneral}

In order to study spontaneous symmetry breaking of time-translation symmetry, we first analyze Floquet systems, which are described by periodic Hamiltonians explicitly breaking this symmetry, $\hat{H}(t+T)=\hat{H}(t)$. The periodic driving may not only be applied in the external potential \cite{Lignier2007,Kierig2008}, but also in the interacting \cite{Gaul2011,Rapp2012} and even in the kinetic term \cite{Stoferle2004,Pieplow2019}. In a many-body system whose dynamics can be described by a variational ansatz of the form (\ref{eq:SelfConsistentVariational}), a periodic Hamiltonian is translated into 
\begin{equation}\label{eq:SelfConsistentVariationalFloquet}
    i\frac{dX^i}{dt}=H^i_{j}(\mathbf{X},t)X^j,
\end{equation}
where the matrix $H^i_{j}(\mathbf{X},t)$ satisfies $H^i_{j}(\mathbf{X},t+T)=H^i_{j}(\mathbf{X},t)$. We assume that the problem admits Floquet solutions of the form
\begin{equation}\label{eq:SelfConsistentFloquetState}
    X^i(t)=X_0^i(t)e^{-i\epsilon^i t},~X_0^i(t+T)=X_0^i(t),
\end{equation}
satisfying the self-consistent Floquet equation 
\begin{equation}\label{eq:SelfConsistentFloquetEquationExternal}
    \epsilon^{i}X_0^i(t)=\left[H^i_{j}(\mathbf{X}_0(t),t)-i\frac{d}{dt}\delta^i_{j}\right]X_0^j(t),
\end{equation}
where $H^i_{j}(\mathbf{X}_0(t),t)$ is now a fully periodic operator as the vector $\mathbf{X}_0(t)$ is also periodic. The dynamics of the collective modes is described by a linear equation analogous to Eq. (\ref{eq:VariationalCollectiveModes}),
\begin{equation}
    i\frac{d \delta X^{i}}{dt}=M^i_{j}(t)\delta X^{j},
\end{equation}
with $M^i_{j}(t+T)=M^i_{j}(t)$. As a linear periodic equation, it admits Floquet solutions of the form $\delta X^{i}(t)=u^i_{\varepsilon}(t)e^{-i\varepsilon t}$, with $u^i_{\varepsilon}(t+T)=u^i_{\varepsilon}(t)$ and $\varepsilon$ the quasi-energy. The corresponding eigenvalue equation reads
\begin{equation}\label{eq:VariationalCollectiveModesFloquet}
    \varepsilon u^i_{\varepsilon}(t)=\left[M^i_{j}(t)-i\frac{d}{dt}\delta^i_{j}\right]u^j_{\varepsilon}(t).
\end{equation} 
Following the reasoning of the stationary case, if the Floquet state (\ref{eq:SelfConsistentFloquetState}) spontaneously breaks a continuous symmetry of the problem, we now have that  
\begin{equation}\label{eq:VariationalGoldstoneFloquet}
    0=\left[M^i_{j}(t)-i\frac{d}{dt}\delta^i_{j}\right] Y^j,
\end{equation}
so $\mathbf{Y}=-iL\mathbf{X}_0$ is a Floquet mode with zero quasi-energy. We can identify it as the NG mode associated to the spontaneous-symmetry breaking, and consequently we will denote it as Floquet-Nambu-Goldstone mode. 

It was recently shown \cite{deNova2022} that, due to many-body interactions, Floquet states can spontaneously emerge even in the absence of external periodic driving as self-consistent Floquet solutions (\ref{eq:SelfConsistentFloquetState}) to the original time-independent problem (\ref{eq:SelfConsistentVariational}), 
\begin{equation}\label{eq:SelfConsistentFloquetEquationSpontaneous}
    \epsilon^{i}X_0^i(t)=\left[H^i_{j}(\mathbf{X}_0(t))-i\frac{d}{dt}\delta^i_{j}\right]X_0^j(t).
\end{equation}
It is important to remark that here the periodicity of $H^i_{j}(\mathbf{X}_0(t))$ stems solely from that of $\mathbf{X}_0(t)$, and is spontaneously determined by the solution itself and not externally imposed, in contrast to Eq. (\ref{eq:SelfConsistentFloquetEquationExternal}). The linear problem is also described by Floquet modes, as in Eq. (\ref{eq:VariationalCollectiveModesFloquet}), but once more the periodicity of the matrix $M^i_{j}(t)$ stems just from that of $\mathbf{X}_0(t)$. Similarly, spontaneous symmetry breaking is translated into the emergence of FNG modes. However, a spontaneous Floquet state also breaks time-translation symmetry since $\mathbf{X}_0(t+t_0)$ is as well a solution of Eq. (\ref{eq:SelfConsistentFloquetEquationSpontaneous}). This implies that $\mathbf{Y}=\partial_t \mathbf{X}_0$ is an FNG mode,
\begin{equation}\label{eq:VariationalGoldstoneFloquetTemporal}
    0=\left[M^i_{j}(t)-i\frac{d}{dt}\delta^i_{j}\right]\partial_t X_0^j,
\end{equation}
a direct consequence of the time-independence of the original Hamiltonian. The presence of a genuine temporal FNG mode is therefore the hallmark of a spontaneous Floquet state, absent in driven Floquet systems or accidental time-periodic states such as that of Eq. (\ref{eq:AccidentalSFstate}).


\section{Simultaneous symmetry breaking in spontaneous Floquet states}\label{sec:SimultaneousSymmetryBreakingFloquet}

We directly adapt the formalism developed in Sec. \ref{sec:SimultaneousSymmetryBreaking} to study simultaneous symmetry breaking in  spontaneous Floquet states, focusing once more on the case of a condensate for simplicity. Within this framework, spontaneous Floquet states are periodic solutions 
\begin{equation}\label{eq:FloquetGPGWaveFunction}
\Psi_0(\mathbf{x},t+T)=\Psi_0(\mathbf{x},t)
\end{equation}
of the constrained Lagrangian (\ref{eq:LagrangianGibbs}), where $\{K,\mathbf{Q}\}$ are still a set of conserved charges. The resulting equation of motion is
\begin{equation}\label{eq:FloquetGPGTime}
    K_{GP} \Psi_0=i\partial_t\Psi_0.
\end{equation}
When inserted into the usual time-dependent GP equation (\ref{eq:GPEquation}), taking $\Psi(\mathbf{x},0)=\Psi_0(\mathbf{x},0)$
yields
\begin{equation}\label{eq:GP2GPGTime}
   \Psi(\mathbf{x},t)=e^{-i\boldsymbol{\lambda}\cdot \mathbf{T} t} \Psi_0(\mathbf{x},t).
\end{equation}
Due to time-translation symmetry, if $\Psi_0(\mathbf{x},t)$ is a solution of this equation, then $\Psi_0(\mathbf{x},t+t_0)$ is also a solution. The structure of these equations suggests to gather hereafter the symmetries $\mathbf{T}$ (which we recall do not involve time by assumption) with the time-translation symmetry as
\begin{equation}\label{eq:RelativisticSymmetryNotation}
    \lambda^\alpha T_\alpha\equiv \boldsymbol{\lambda}\cdot \mathbf{T}+i\partial_t.
\end{equation}
This addition also reflects the emergence of an extra NG mode, $z_t$, associated to the time-translation symmetry breaking, whose conserved charge is in turn $Q_t=K$. In particular, the Goldstone-Gibbs modes $z_a(t)$ are now periodic, $z_a(t+T)=z_a(t)$, satisfying
\begin{equation}\label{eq:GoldstoneGibbsFloquetTime}
    [M_0(t)-i\partial_t]z_\alpha=0,~[M_0(t)-i\partial_t]z_A=i(\partial_A\lambda^\beta) z_\beta,
\end{equation} 
where $M_0(t)$ is the periodic operator resulting from the substitution $\Psi_0(\mathbf{x})\to \Psi_0(\mathbf{x},t)$ in Eq. (\ref{eq:BdGGMatrix}).
Thus, the modes $z_\alpha(t)$ are zero quasi-energy modes, which we can identify as the FNG modes associated to the spontaneous symmetry breaking by virtue of Eq. (\ref{eq:VariationalGoldstoneFloquet}). The Floquet spectrum is completed by the regular Bogoliubov sector 
\begin{equation}\label{eq:BogoliubovFloquetSectorTime}
    [M_0(t)-i\partial_t]z_{n}=\varepsilon_n z_{n},
\end{equation}
with $z_{n}(\mathbf{x},t+T)=z_{n}(\mathbf{x},t)$ and $\varepsilon_n$ the quasi-energy (defined modulo $\omega=2\pi/T$). With the help from the conventional Floquet theory and from the results of Secs. \ref{subsec:BdGG}, \ref{subsec:BerryGibbs}, it is shown that the orthogonality relations (\ref{eq:OrthogonalityBdGG}) still hold:
\begin{equation}\label{eq:OrthogonalityBdGGFloquet}
(z_{n}|z_{n'})=\delta_{nn'},~(z_a|z_{n})=0,~(z_a|z_b)=i\Omega_{ab}.
\end{equation}
The dynamics of the field operator is again described by the time-independent Heisenberg equation of motion (\ref{eq:HeisenbergEquationOfMotionBoson}). However, when working in the Gibbs picture (\ref{eq:GibbsPicture}), we now expand the field operator around the wavefunction of the spontaneous Floquet state, $\hat{\Psi}'(\mathbf{x},t)=\Psi_0(\mathbf{x},t)+\hat{\varphi}(\mathbf{x},t)$, arriving at the time-dependent BdGG equations $i\partial_t \hat{\Phi}=M_0(t)\hat{\Phi}$.
Using the complete Floquet spectrum, we can express the quantum fluctuations of the field operator as
\begin{align}\label{eq:FloquetFieldFluctuations}
  \nonumber &\hat{\Phi}(\mathbf{x},t)=\hat{X}^\alpha(t) z_\alpha(\mathbf{x},t)+\hat{P}^A(t)z_A(\mathbf{x},t)\\
  &+\sum_{n}\hat{\gamma}_{n}(t)z_{n}(\mathbf{x},t)+\hat{\gamma}^{\dagger}_{n}(t)\bar{z}_{n}(\mathbf{x},t).
\end{align}

In general, for any time-dependent spinor $z(\mathbf{x},t)$, we can define an  amplitude operator as $\hat{\gamma}(t)\equiv (z(t)|\hat{\Phi}(t))$, satisfying $[\hat{\gamma},\hat{\gamma}'^\dagger]=(z(t)|z'(t))$, with $\hat{\gamma}'(t)=(z'(t)|\hat{\Phi}(t))$.
After taking into account the time dependence of $z(\mathbf{x},t)$, the equation of motion for $\hat{\gamma}(t)$ is simply 
\begin{eqnarray}
i\partial_t \hat{\gamma}(t)=([M_0(t)-i\partial_t]z|\hat{\Phi})
\end{eqnarray}
Applying this relation to each mode results in the same formal time dependence for the quantum amplitudes as in the stationary case, Eq. (\ref{eq:QuantumFieldFluctuationsExplicit}), namely:
\begin{eqnarray}\label{eq:QuantumFieldFluctuationsExplicitTimeFloquet}
\nonumber \hat{\gamma}_{n}(t)&=&\hat{\gamma}_{n}e^{-i\varepsilon_n t},\\
\hat{P}^A(t)&=&\hat{P}^A,\\
\nonumber \hat{X}^\alpha(t)&=&\hat{X}^\alpha+\partial_A\lambda_B \hat{P}^B t.
\end{eqnarray}
The problem of this formalism is that the role of the Lagrange multiplier associated with time is not clear since $\lambda^t=1$ is fixed by the Schr\"odinger equation. This is also seen from the fact that the associated charge is the generalized Gibbs Hamiltonian $K$, which should naively satisfy $\partial_K E=1$. The corresponding momentum $\hat{P}^K=-i(z_t|\hat{\Phi})$ describes the quantum fluctuations of $K$,
\begin{equation}
    \hat{K}=\hat{H}-\boldsymbol{\lambda}\cdot\mathbf{\hat{Q}}\simeq  K+\hat{P}^K.
\end{equation}
However, since the preservation of the commutation relations implies $\partial_A\lambda^t=0=\partial_K\lambda^\alpha$, Eq. (\ref{eq:CommutationPreservation}), the momentum $\hat{P}^K$ does not lead to any dynamics, which is contradictory as fluctuations in $K$ should indeed affect the system. Moreover, we lack a satisfying thermodynamical description because now the energy is also an independent variable. All of this suggests that we need to extend our formalism to attain the complete picture. 


\begin{figure}[tb!]
    \includegraphics[width=\columnwidth]{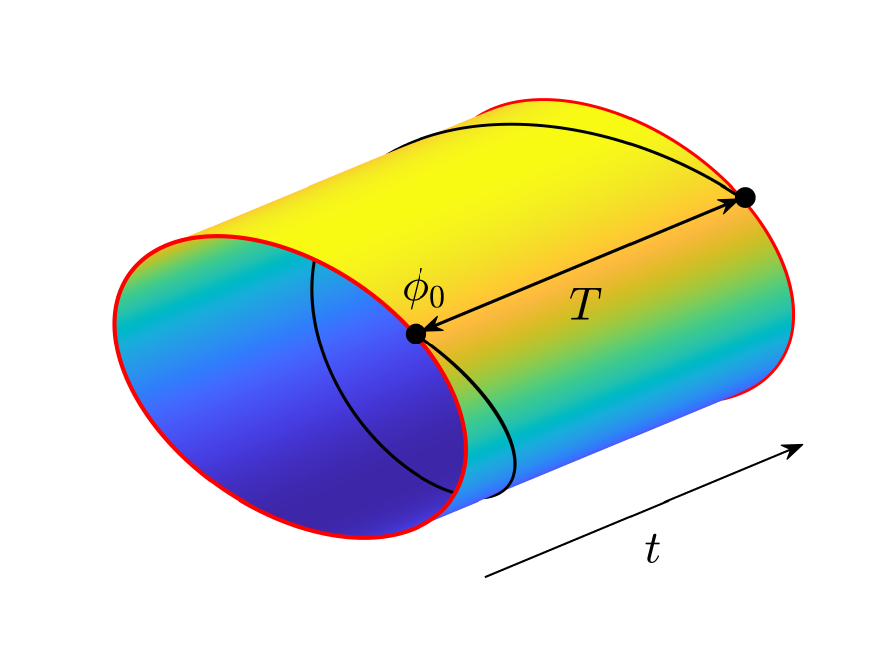} 
\caption{Schematic representation of the $(t,\phi)$ formalism, where $t$ parametrizes the longitudinal axis of the cylinder and $\phi$ its polar angle. The dynamics for the field $\Psi(\mathbf{x},\phi,t)$ is solved over the whole cylinder surface while the physical field $\Psi(\mathbf{x},t)=\Psi(\mathbf{x},\phi_0+\omega t,t)$ is restricted to the spiral trajectory given by the solid black line. The red circles correspond to the constant sections $t=0$ and $t=T$.} 
\label{fig:Cilindro}
\end{figure}

\subsection{$(t,\phi)$ formalism}\label{subsec:tefi}

In order to correctly derive the dynamics of the temporal FNG mode, we develop the $(t,\phi)$ formalism, denoted in this way due to its analogy with the more usual $(t,t')$ formalism \cite{Peskin1993,Heimsoth2012}. Here, $\phi\in [0,2\pi)$ is the angular variable associated to the periodic motion of the spontaneous Floquet state, $\phi=\omega t'$, and $t$ is a proper time that accounts for the nonperiodic dynamics. As we will see, the use of $\phi$ instead of $t'$ is critical. 

In the $(t,\phi)$ formalism, we add an extra dimension to the field, $\Psi=\Psi(\mathbf{x},\phi,t)$, whose dynamics is described by the Lagrangian 
\begin{equation}\label{eq:LagrangianTefi} L=\frac{1}{2\pi}\int^{2\pi}_0\mathrm{d}\phi\left(\int\mathrm{d}\mathbf{x}~[i\Psi^\dagger \partial_t\Psi+i\Psi^\dagger\omega \partial_\phi\Psi]-H\right),
\end{equation}
where $H$ does not explicitly depend on $\phi$ as the Hamiltonian is time-independent. The case of a conventional, driven Floquet system is simply accounted for by a $\phi$-dependent periodic Hamiltonian, $H(\phi+2\pi)=H(\phi)$. 

The resulting equation of motion for $\Psi(\mathbf{x},\phi,t)$ is the classical version of Eq. (\ref{eq:HeisenbergEquationOfMotion}) with the substitution 
\begin{equation}\label{eq:FloquetSub}
    i\partial_t\Psi(\mathbf{x},t)\to i\omega \partial_\phi\Psi(\mathbf{x},\phi,t)+i \partial_t\Psi(\mathbf{x},\phi,t).
\end{equation}
The dynamics of the field, accounted by the term $i\partial_t\Psi(\mathbf{x},t)$, is then split into the periodic contribution (either determined by the periodicity of the spontaneous Floquet state or by the external driving), parametrized by the phase $\phi$ and encoded in the term $i\omega \partial_\phi\Psi(\mathbf{x},\phi,t)$, and the nonperiodic contribution, parametrized by the time $t$ and encoded in the term $i \partial_t\Psi(\mathbf{x},\phi,t)$. Once the dynamics is solved for $\Psi(\mathbf{x},\phi,t)$, the physical field is obtained as $\Psi(\mathbf{x},t)=\Psi(\mathbf{x},\phi_0+\omega t,t)$, with $\phi_0$ some initial phase. When the system is driven, this phase is locked by the actual time-dependence of the Hamiltonian, $H(t)=H(\phi_0+\omega t)$, which we can conventionally set to $\phi_0=0$. However, for spontaneous Floquet states, $\phi_0$ is a free parameter, reflecting the time-independence of the original Hamiltonian. 

Intuitively, we can visualize the $(t,\phi)$ space as a cylinder where $t$ parametrizes its longitudinal axis and $\phi$ its polar angle, Fig. \ref{fig:Cilindro}. The field $\Psi(\mathbf{x},\phi,t)$ is defined over the whole surface of the cylinder, while the actual physical field $\Psi(\mathbf{x},t)$ is evaluated solely along the spiral trajectory $(t,\phi_0+\omega t)$ (solid black line). 

Regarding the conserved charges, the action is the time integral of the Lagrangian (\ref{eq:LagrangianTefi}), to which we apply Noether theorem. The extra variable $\phi$ simply adds an angular average to Eq. (\ref{eq:ClassicalNoetherCharge}). In particular, the usual charges $\mathbf{Q}$ arising from the continuous symmetries of the Hamiltonian are readily retrieved as
\begin{equation}
\mathbf{Q}=\frac{1}{2\pi}\int^{2\pi}_0\mathrm{d}\phi~ \mathbf{Q}(\phi)= \braket{\Psi||\mathbf{T}||\Psi},
\end{equation}
with
\begin{eqnarray}\label{eq:ExtendendScalarProduct}
    \nonumber \braket{\chi||\Psi}&\equiv& \frac{1}{2\pi}\int^{2\pi}_0\mathrm{d}\phi\int\mathrm{d}\mathbf{x}~\chi^\dagger(\mathbf{x},\phi)\Psi(\mathbf{x},\phi)\\
    &=&\frac{1}{2\pi}\int^{2\pi}_0\mathrm{d}\phi~\braket{\chi(\phi)|\Psi(\phi)}
\end{eqnarray}
the scalar product in the extended Hilbert space.


In addition, we now have an extra conserved charge $Q_\phi=F$, which we will denote as the Floquet charge, resulting from invariance under phase translations $\phi\to \phi+\phi_0$:
\begin{equation}
    F=\frac{1}{2\pi}\int^{2\pi}_0\mathrm{d}\phi\int\mathrm{d}\mathbf{x}~i\Psi^\dagger\partial_\phi\Psi=\braket{\Psi||T_\phi||\Psi},
\end{equation}
with $T_\phi=i\partial_\phi$ the generator of phase translations. The operator $T_\phi$ is Hermitian under the extended scalar product (\ref{eq:ExtendendScalarProduct}), and commutes with the rest of generators, $[T_\phi,\mathbf{T}]=0$. 

Finally, invariance under time translations $t\to t+t_0$ results in the conservation of the Floquet enthalpy [see Eq. (\ref{eq:FloquetEnthalpyConservation}) and ensuing discussion] 
\begin{equation}\label{eq:FloquetEnergapy}
    I\equiv E-\omega F.
\end{equation}
Nevertheless, while $I$ is conserved for both driven and spontaneous Floquet states, the conservation of the Floquet charge $F$ is a genuine feature of a spontaneous Floquet state since the Hamiltonian is $\phi$-dependent for driven Floquet systems. 

We note that the $(t,\phi)$ formalism can be applied to spontaneous and conventional Floquet systems in any many-body system governed by a second-quantization Hamiltonian of the general form (\ref{eq:HamiltonianManyBody}), including both bosons and fermions, since the dynamics of the field operator then admits a Lagrangian description, as discussed at the beginning of Sec. \ref{sec:SimultaneousSymmetryBreaking}.







\subsection{Spontaneous Floquet states in the generalized Gibbs ensemble: Floquet thermodynamics}\label{subsec:FloquetGPG}

We invoke again the generalized Gibbs ensemble to describe spontaneous symmetry breaking within the $(t,\phi)$ formalism, replacing the Hamiltonian $H$ by the generalized Gibbs Hamiltonian $K$ in Eq. (\ref{eq:LagrangianTefi}), in analogy with the constrained Lagrangian  (\ref{eq:LagrangianGibbs}). As a result, the Floquet enthalpy $I$ is replaced by the generalized Floquet-Gibbs energy   
\begin{equation}
    \Lambda\equiv K-\omega F=I-\boldsymbol{\lambda}\cdot \mathbf{Q}=H-\lambda^\alpha Q_\alpha,
\end{equation}
which is the analogue of the generalized Gibbs energy for Floquet systems. The expression $\lambda^\alpha Q_\alpha\equiv \boldsymbol{\lambda}\cdot \mathbf{Q}+\omega F$  allows us to identify $\lambda^\phi=\omega$ as the Lagrange multiplier of the Floquet charge.

In the case of a condensate, the generalized Floquet-Gibbs energy $\Lambda$ is the thermodynamic potential that is extremized by stationary solutions $\Psi_0(\mathbf{x},\phi)$, satisfying the Floquet-GPG equation 
\begin{equation}
    0=\frac{\delta \Lambda}{\delta \Psi^*}\Longrightarrow\Lambda_{GP}\Psi_0\equiv [K_{GP}-i\omega\partial_\phi]\Psi_0=0.
\end{equation}
When we insert this wavefunction into the actual time-dependent GP equation (\ref{eq:GPEquation}) by taking $\Psi(\mathbf{x},t=0)=\Psi_0(\mathbf{x},\phi_0)
$, we find
\begin{eqnarray}\label{eq:UnitaryTefi}
   \Psi(\mathbf{x},t)&=&e^{-i\lambda^\alpha T_\alpha t}\Psi_0(\mathbf{x},\phi_0)\\
   \nonumber&=&e^{-i\boldsymbol{\lambda}\cdot \mathbf{T} t} \Psi_0(\mathbf{x},\phi_0+\omega t).
\end{eqnarray} 
Thus, Floquet states are described by stationary solutions with $\phi=\phi_0$, which behaves as a global time origin $t_0=\phi_0/\omega$ in the actual GP solution. In a spontaneous Floquet state, this time origin can be chosen freely due to the time-independence of the Hamiltonian. In a conventional Floquet state, $t_0$ is fixed by the external driving, as discussed after Eq. (\ref{eq:FloquetSub}). On the other hand, as expected from thermodynamic considerations, $\omega$ plays the predicted role for the Lagrange multiplier associated to the Floquet charge, acting as the velocity of phase translations. In fact, the definition modulo $\omega$ of the chemical potential $\mu$ for Floquet states can be also explained from Thermodynamics as follows. The transformation $\Psi_0(\mathbf{x},\phi)\to \Psi_0(\mathbf{x},\phi)e^{-in\phi}$, $n\in\mathbb{Z}$, also yields a solution of the Floquet-GPG equation. This leads to $F\to F+n N$, so $\Lambda\to \Lambda -n\omega N$, and hence $\lambda^\theta=\mu$ changes as $\mu \to \mu+n\omega$.


Another hallmark of spontaneous Floquet states is that they conserve energy, $E=\Lambda+\lambda^\alpha Q_\alpha$, since $\{\Lambda,Q_\alpha\}$ are conserved charges; the argument fails for driven Floquet systems because there $F$ is not conserved. Following Eqs. (\ref{eq:EExpansion})-(\ref{eq:GibbsDuhem}), it can be shown that each Lagrange multiplier can be obtained from the energy as $\partial_A E=\lambda_A$, including $\partial_F E=\lambda_F=\omega$. Moreover, if we now define the local pressure $p(\mathbf{x})$ as the angular average
\begin{equation}
    p(\mathbf{x})\equiv \frac{1}{2\pi}\int^{2\pi}_0\mathrm{d}\phi~\frac{|\Psi_0(\mathbf{x},\phi)|^4}{2},
\end{equation}
we retrieve the generalization of the first principle of Thermodynamics 
\begin{equation}\label{eq:FirstTermoquet}
    dE=\lambda^\alpha dQ_\alpha- pdV=\boldsymbol{\lambda}\cdot d\mathbf{Q}+\omega dF-pdV,
\end{equation}
and of the Gibbs-Duhem relation (\ref{eq:GibbsDuhem}) for spontaneous Floquet states. 

The upshot of the above discussion is that Floquet states allow for a thermodynamical description completely analogous to that of stationary states, which we denote as Floquet thermodynamics. Indeed, the first principle of Thermodynamics for spontaneous Floquet states (\ref{eq:FirstTermoquet}) suggests a fundamental distinction between spontaneous and driven Floquet systems: spontaneous systems conserve the charge $F$ and the frequency $\omega$ is determined by the equation of state, while driven systems externally impose the frequency $\omega$ so then $F$ is not conserved. Thus, we can regard spontaneous Floquet systems as ``isofloquetic'', where the macroscopic Floquet charge is conserved, in analogy with isentropic or isochoric systems, where the entropy or the volume is conserved. Conversely, conventional Floquet systems are ``isoperiodic'', where the external driving fixes the oscillation period, in analogy with isothermal or isobaric systems, where the external environment fixes the temperature or the pressure. Based on this analogy, $I=E-\omega F$ can be identified as the Floquet enthalpy, the natural thermodynamic potential for isoperiodic systems, where it is conserved and satisfies
\begin{equation}\label{eq:FloquetEnthalpyConservation}
     dI=\boldsymbol{\lambda}\cdot d\mathbf{Q}-Fd\omega -pdV.
\end{equation}
Thus, we can also provide a thermodynamical description for conventional Floquet states by using the Floquet enthalpy, from where the Lagrange multipliers are simply derived as $\lambda_A=\partial_A I$.

Physically, the Floquet charge accounts for the independence of the energy, allowing for the emergence of the temporal FNG mode. This is because the stationarity condition for solutions of the Lagrangian (\ref{eq:LagrangianGibbs}) imposes a constraint from where the energy becomes a function of the remaining charges, $E=E(\mathbf{Q})$. As argued at the end of Sec. \ref{sec:Cnoidal}, this dependence implies that time-translation symmetry is not spontaneously broken by itself but rather in terms of other symmetries. However, for spontaneous Floquet states, energy becomes an independent charge, and the thermodynamic magnitude whose natural variables are the energy and the other conserved charges is precisely the Floquet charge, $F=F(E,\mathbf{Q})$. This is translated into the emergence of a genuine, independent temporal FNG mode. 


\subsection{Quantization}\label{subsec:QuantumFloquet}

The $(t,\phi)$ formalism arises in the quantum theory when considering once more the time-independent Heisenberg equation (\ref{eq:HeisenbergEquationOfMotionBoson}). In order to work in the rotating frame, we perform a unitary transformation to the field operator  along the lines of Eq. (\ref{eq:UnitaryTefi}),
\begin{widetext}
\begin{equation}
    \hat{\Psi}(\mathbf{x},t)=e^{-i\boldsymbol{\lambda}\cdot \mathbf{T} t} \hat{\Psi}(\mathbf{x},\phi+\omega t,t)=e^{-i\lambda^\alpha T_\alpha t}\hat{\Psi}(\mathbf{x},\phi,t).
\end{equation}
In the specific case of a condensate, the resulting equation of motion for the field operator in the $(t,\phi)$ formalism reads

\begin{align}\label{eq:HeisenbergEquationOfMotionTeFi}
    i\hbar\partial_t\hat{\Psi}(\mathbf{x},\phi,t)=\left[-\frac{\nabla^2}{2}+V(\mathbf{x})+\hat{\Psi}^{\dagger}(\mathbf{x},\phi,t)\hat{\Psi}(\mathbf{x},\phi,t)-\lambda^\alpha T_\alpha\right]\hat{\Psi}(\mathbf{x},\phi,t),
\end{align}
\end{widetext}
which is the quantum version of the time-dependent Floquet-GPG equation. A similar equation can be derived for a conventional Floquet system, where the frequency is fixed by the external driving provided by the periodic Hamiltonian $\hat{H}(\phi)$. Once the dynamics in the $(t,\phi)$ formalism is computed, the actual field operator is retrieved as $\hat{\Psi}(\mathbf{x},t)=e^{-i\boldsymbol{\lambda}\cdot \mathbf{T} t}\hat{\Psi}(\mathbf{x},\phi_0+\omega t,t)$.

After expanding the field operator around the stationary Floquet state, $\hat{\Psi}(\mathbf{x},\phi,t)=\Psi_0(\mathbf{x},\phi)+\hat{\varphi}(\mathbf{x},\phi,t)$, we obtain at linear order from Eq. (\ref{eq:HeisenbergEquationOfMotionTeFi}) that
\begin{eqnarray}
\nonumber i\partial_t \hat{\Phi}&=&[M_0(\phi)-i\omega\partial_\phi]\hat{\Phi},\\
\hat{\Phi}(\mathbf{x},\phi,t)&=&\left[\begin{array}{c}\hat{\varphi}(\mathbf{x},\phi,t)\\ \hat{\varphi}^{\dagger}(\mathbf{x},\phi,t)\end{array}\right].
\end{eqnarray}
The Bogoliubov modes are derived in the same fashion of Eqs. (\ref{eq:GoldstoneGibbsFloquetTime}), (\ref{eq:BogoliubovFloquetSectorTime}), replacing $M_0(t)-i\partial_t$ by $M_0(\phi)-i\omega\partial_\phi$ and using the extended inner product 
\begin{equation}\label{eq:KGProduct}
(z||z')\equiv\braket{z||\sigma_z z'}=\frac{1}{2\pi}\int^{2\pi}_0\mathrm{d}\phi~(z(\phi)|z'(\phi)).
\end{equation}
The Goldstone-Gibbs modes $z_a(\mathbf{x},\phi)$ are given here in terms of $\partial_a \Psi_0(\mathbf{x},\phi)$, while the Bogoliubov sector $z_{n}(\mathbf{x},\phi)$ is obtained by the adaptation of the Floquet modes of Eq. (\ref{eq:BogoliubovFloquetSectorTime}), $z_{n}(\mathbf{x},t)\equiv z_{n}(\mathbf{x},\phi_0+\omega t)$. In order to have a complete orthonormal basis in the extended Hilbert space, we promote each spinor $z(\mathbf{x},\phi)$ to a whole discrete set 
\begin{equation}
    z_{m}(\mathbf{x},\phi)\equiv z(\mathbf{x},\phi)e^{im\phi},~m\in\mathbb{Z}.
\end{equation}

Following the prescriptions of Secs. \ref{subsec:BdGG}, \ref{subsec:BerryGibbs}, it is immediate to show that these modes satisfy
\begin{align}
      &[M_0(\phi)-i\omega\partial_\phi]z_{\alpha,m}=m\omega z_{\alpha,m},\\
      &\nonumber [M_0(\phi)-i\omega\partial_\phi]z_{A,m}=m\omega z_{A,m}+i\partial_A\lambda^\beta z_{\beta,m},\\
       \nonumber &[M_0(\phi)-i\omega\partial_\phi]z_{n,m}=[\varepsilon_n+m\omega] z_{n,m},
\end{align}
and do form a complete orthonormal basis, obeying the correct orthogonality relations
\begin{eqnarray}
    (z_{n,m}||z_{n',m'})&=&\delta_{nn'}\delta_{mm'},\\
    \nonumber (z_{a,m}||z_{n,m'})&=&0,\\
    \nonumber (z_{a,m}||z_{b,m'})&=&i\Omega_{ab}\delta_{mm'}.
\end{eqnarray}
Notice that here the Berry-Gibbs curvature (\ref{eq:BerryGibbsCurvature}) is evaluated using the extended scalar product (\ref{eq:ExtendendScalarProduct}).
\begin{widetext}

Quantum amplitudes are now obtained by $\hat{\gamma}_{m}(t)=(z_{m}||\hat{\Phi}(t))$. From the Fourier relations \\ 
\begin{eqnarray} 
\hat{\gamma}_{m}(t)&=&\frac{1}{2\pi}\int^{2\pi}_0\mathrm{d}\phi~e^{-im\phi}\hat{\gamma}(\phi,t),\\
     \nonumber \hat{\gamma}(\phi,t)&\equiv &(z(\phi)|\hat{\Phi}(\phi,t))=\sum_{m=-\infty}^\infty \hat{\gamma}_{m}(t)e^{im\phi},
\end{eqnarray}
we can expand the quantum fluctuations of the field operator as

\begin{align}\label{eq:FloquetPhaseFieldFluctuations}
   \nonumber \hat{\Phi}(\mathbf{x},\phi,t)&=\sum_{m=-\infty}^\infty\hat{X}_m^\alpha(t) z_{\alpha,m}(\mathbf{x},\phi)+\hat{P}_m^A(t)z_{A,m}(\mathbf{x},\phi)+\sum_{m=-\infty}^\infty\sum_{n}\hat{\gamma}_{n,m}(t)z_{n,m}(\mathbf{x},\phi)+\hat{\gamma}^{\dagger}_{n.m}(t)\bar{z}_{n,m}(\mathbf{x},\phi)
   \\ &=\hat{X}^\alpha(\phi,t) z_{\alpha}(\mathbf{x},\phi)+\hat{P}^A(\phi,t)z_{A}(\mathbf{x},\phi)+\sum_{n}\hat{\gamma}_{n}(\phi,t)z_{n}(\mathbf{x},\phi)+\hat{\gamma}^{\dagger}_{n}(\phi,t)\bar{z}_{n}(\mathbf{x},\phi).
\end{align}
The dynamics for $\hat{\Phi}(\mathbf{x},\phi,t)$ is readily obtained by the relation $i\partial_t\hat{\gamma}_{m}(t)=([M_0(\phi)-i\omega\partial_\phi]z_{m}||\hat{\Phi}(t))$, finding:
\begin{equation}
     \hat{\gamma}_{n,m}(t)=\hat{\gamma}_{n,m}e^{-i[\varepsilon_n+m\omega]t},~~~
    \hat{P}_m^A(t)=\hat{P}_m^Ae^{-im\omega t},~~~\hat{X}_m^\alpha(t)=[\hat{X}_m^\alpha+\partial_A\lambda_B\hat{P}_m^B t]e^{-im\omega t}.
\end{equation}
Inverting the Fourier transformation yields:
\begin{equation}
     \hat{\gamma}_{n}(\phi,t)=\hat{\gamma}_{n}(\phi-\omega t,0)e^{-i\varepsilon_nt},~~~
    \hat{P}^A(\phi,t)=\hat{P}^A(\phi-\omega t,0),~~~\hat{X}^\alpha(\phi,t)=\hat{X}^\alpha(\phi-\omega t,0)+\partial_A\lambda_B\hat{P}^B(\phi-\omega t,0) t.
\end{equation}

When setting $\phi=\phi_0+\omega t$, after noticing that the standard time-dependent spinors and amplitudes are $z(\mathbf{x},t)=z(\mathbf{x},\phi_0+\omega t)$ and $\hat{\gamma}(t)=\hat{\gamma}(\phi_0+\omega t,t)$, with $\hat{\gamma}\equiv \hat{\gamma}(t=0)=\hat{\gamma}(\phi_0,0)$ their initial values, we recover the expected time evolution for the field fluctuations:
\begin{equation}\label{eq:FloquetFieldPhiFluctuationsTimeDependent}
     \hat{\Phi}(\mathbf{x},t)=\hat{\Phi}(\mathbf{x},\phi_0+\omega t,t)=[\hat{X}^\alpha+\partial_A\lambda_B\hat{P}^B t ]z_{\alpha}(\mathbf{x},t)+\hat{P}^A z_{A}(\mathbf{x},t)+\sum_{n}\hat{\gamma}_{n}z_{n}(\mathbf{x},t)e^{-i\varepsilon_nt}+\hat{\gamma}^{\dagger}_{n}\bar{z}_{n}(\mathbf{x},t)e^{i\varepsilon_nt}.
\end{equation}
\end{widetext}
Formally, this is the same expansion of Eqs. (\ref{eq:FloquetFieldFluctuations}), (\ref{eq:QuantumFieldFluctuationsExplicitTimeFloquet}), which in turn presented the same structure of the stationary case, Eq. (\ref{eq:QuantumFieldFluctuationsExplicit}). The crucial difference is that now we use $(\phi,\omega,F)$ instead of $(t,1,K)$, so the Lagrange multiplier $\lambda^\phi=\omega=\partial_F E$ does lead to the correct (thermo)dynamics. The failure of the original time-dependent approach is that the Gibbs modes there encode a non-trivial dependence in $z_\phi$ since 
\begin{eqnarray}
\partial_A\Psi_0(\mathbf{x},t)&=&\partial_A\Psi_0(\mathbf{x},\phi_0+\omega t)\\
\nonumber &+&\partial_\phi \Psi_0(\mathbf{x},\phi_0+\omega t) \partial_A\omega t.
\end{eqnarray}
When this dependence is taken into account, as well as the thermodynamic relation between $K$ and $F$, after some tedious algebra one arrives at the correct expansion (\ref{eq:FloquetFieldPhiFluctuationsTimeDependent}) from the original Floquet modes of Eqs. (\ref{eq:GoldstoneGibbsFloquetTime}), (\ref{eq:BogoliubovFloquetSectorTime}). 

We note that the above expansion is also valid for driven Floquet systems by excluding the Goldstone-Gibbs pair $\{z_\phi,z_F\}$ and using $\lambda_A=\partial_A I$ instead of $\lambda_A=\partial_A E$. 

In summary, the $(t,\phi)$ formalism allows to treat Floquet states as stationary states by replacing time-independent wavefunctions by periodic wavefunctions, the generalized Gibbs energy $K$ by the generalized Floquet-Gibbs energy $\Lambda=K-\omega F$, energies $\epsilon$ by quasi-energies $\varepsilon$, and conserved charges by their angular averages:
\begin{eqnarray}
\Psi_0(\mathbf{x}),z(\mathbf{x})&\longrightarrow& \Psi_0(\mathbf{x},\phi),z(\mathbf{x},\phi),\\ \nonumber K&\longrightarrow& \Lambda,\\
    \nonumber \epsilon &\longrightarrow&  \varepsilon,\\
    \nonumber Q_\alpha &\longrightarrow&  \frac{1}{2\pi}\int^{2\pi}_0\mathrm{d}\phi~Q_\alpha(\phi).
\end{eqnarray}
For spontaneous Floquet states, one also needs to add the Floquet charge $F$, with no stationary counterpart, while for driven Floquet states the energy $E$ is replaced by the Floquet enthalpy $I$ as the correct thermodynamic potential.

\subsection{Time operator}\label{subsec:TimeOperator}

The previous results demonstrate that the $(t,\phi)$ formalism is not a mere reparametrization of the periodic time $t'$ from the $(t,t')$ formalism, but that it is instead essential to correctly identify the thermodynamical role of the Floquet charge $F=Q_\phi$ and to compute the dynamics of the temporal FNG mode $z_\phi$. In particular, we can explicitly write the time evolution of its quantum amplitude,
\begin{equation}\label{eq:TimeOperatorEvolution}
\hat{X}^\phi(t)=\hat{X}^\phi+\partial_F\lambda_A \hat{P}^A t=\hat{X}^\phi+\partial_A\omega\hat{P}^A t.
\end{equation}
The linear time dependence arises due to the quantum fluctuations of the frequency since this depends on the conserved charges, in analogy to the stationary case [see Eq. (\ref{eq:LagrangeQF}) and ensuing discussion]. The actual temporal FNG mode, corresponding to proper time translations, is simply obtained as $z_t=\omega z_\phi$, with an amplitude $\hat{X}^t=\hat{X}^\phi/\omega$.

On the other hand, the quantum fluctuations $\delta\hat{F}$ of the Floquet charge are proportional to those of the more physical generalized Gibbs energy,
\begin{equation}
\delta\hat{K}=\hat{P}^K=\omega\delta\hat{F}=\omega \hat{P}^F.
\end{equation}
This relation stems from the fact that spontaneous Floquet states extremize the Floquet-Gibbs energy $\Lambda$, so $0=\delta \Lambda=\delta K-\omega \delta F$. As a result, we have the commutation relation
\begin{equation}
    [\hat{X}^\phi,\hat{P}^F]=[\hat{X}^t,\hat{P}^K]=i,
\end{equation}
which shows that $\delta\hat{t}\equiv -\hat{X}^t$ behaves as an effective time operator in the Gibbs picture, describing the quantum fluctuations of the global time shift $t_0$ of the spontaneous Floquet state. 

In general, no time operator $\hat{T}$ is allowed in Quantum Mechanics since the generator of time translations is the Hamiltonian itself,  $[\hat{T},\hat{H}]=-i$, so
\begin{equation}
    e^{-i\hat{T}E_0}\hat{H}e^{i\hat{T}E_0}=\hat{H}-E_0
\end{equation}
for arbitrary value of $E_0$. This implies that, if $E$ is an eigenenergy of the Hamiltonian, then $E-E_0$ is also an eigenergy. However, Hamiltonians are bounded from below by the ground-state energy, leading to a contradiction.

This relation is very similar to that between the particle number and phase operators, which we review following the enlightening discussion of Ref. \cite{Mora2003}. If a phase operator $\hat{\theta}$ exists, then it satisfies $[\hat{N},\hat{\theta}]=i$. But
\begin{equation}
    e^{-i\hat{\theta}N_0}\hat{N}e^{i\hat{\theta}N_0}=\hat{N}-N_0
\end{equation}
implies that, if $N$ is an eigenvalue of $\hat{N}$, then $N-N_0$ is also. This violates two fundamental properties of the spectrum of the particle-number operator: its positive definiteness and its discreteness. Nevertheless, one can define phase fluctuations in situations involving large particle numbers, where one can neglect the discreteness of the spectrum and states with low occupation number, as in the case of a condensate. Analogously, by identifying energy with particle number and time with phase, one can define time fluctuations for large energies, well above the ground state. Actually, within the $(t,\phi)$ formalism, the time operator is a sort of phase operator. This analogy suggests that spontaneous Floquet states should be highly excited states, shifted by a macroscopic energy from the ground state. 

\section{Time supersolids: CES state}\label{sec:ces}

\begin{figure*}[!tb]

\begin{tabular}{@{}cc@{}}
     \stackinset{l}{0pt}{t}{0pt}{\textbf{(a)}}{\includegraphics[width=\columnwidth]{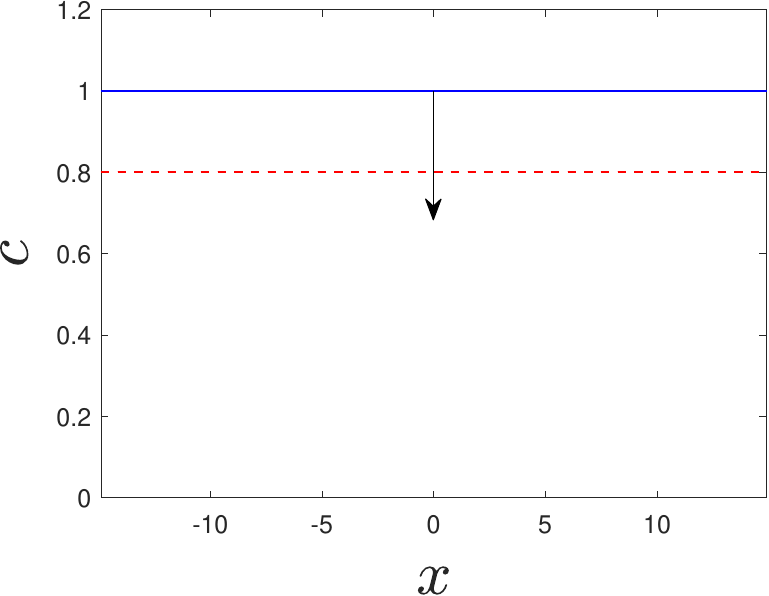}} ~~&~~ 
     \stackinset{l}{0pt}{t}{0pt}{\textbf{(b)}}{\includegraphics[width=\columnwidth]{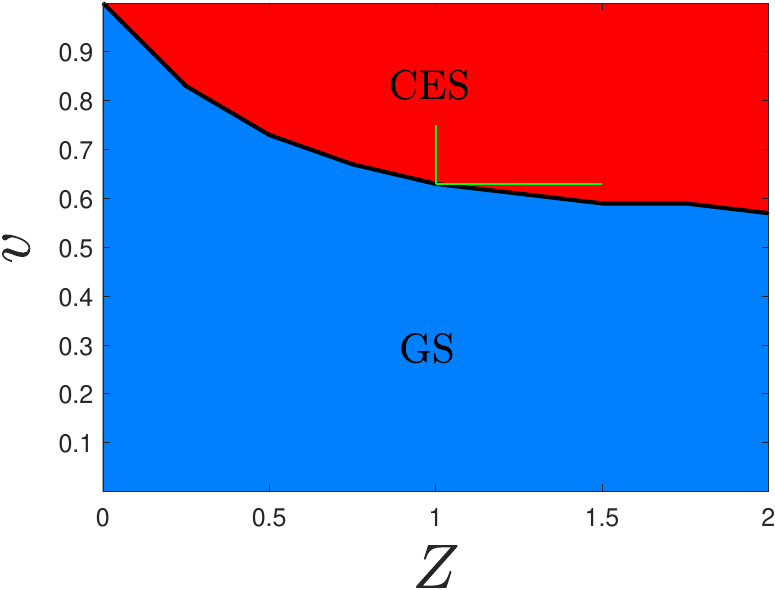}} \\ \\ \\
     \stackinset{l}{0pt}{t}{0pt}{\textbf{(c)}}{\includegraphics[width=\columnwidth]{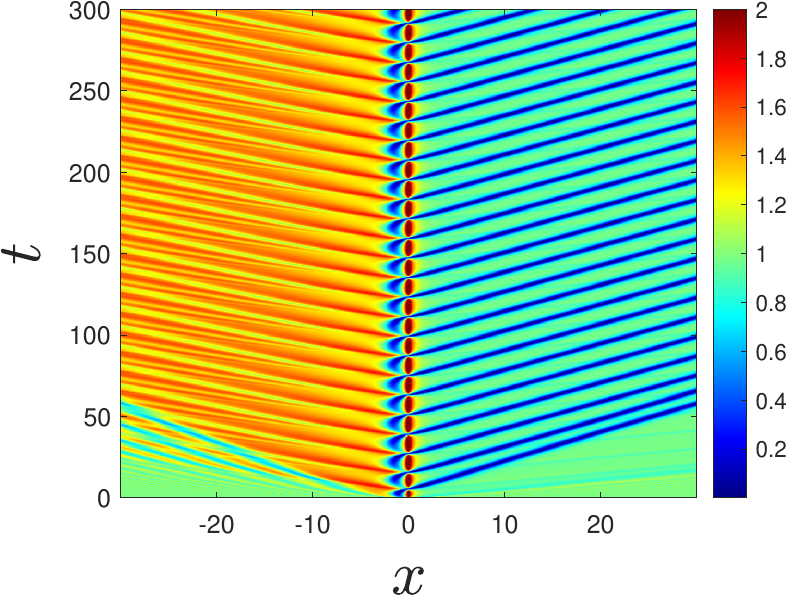}} ~~&~~  
     \stackinset{l}{0pt}{t}{0pt}{\textbf{(d)}}{\includegraphics[width=\columnwidth]{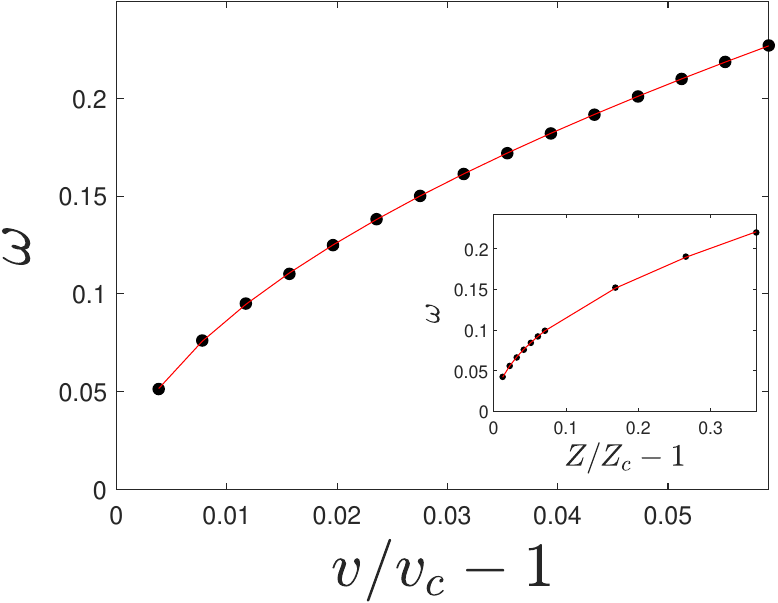}}
\end{tabular}
\caption{(a) Schematic profile of the sound (solid blue) and flow (dashed red) velocities of the initial homogeneous flowing condensate, $\Psi(x,0)=e^{ivx}$. The arrow represents the attractive delta potential $V(x)=-Z\delta(x)$, suddenly introduced at $t=0$. (b) Dynamical phase diagram for the final state of the system as a function of $(Z,v)$. (c) 2D plot of $|\Psi(x,t)|^2$ resulting from the evolution of (a) for $v=0.8$ and $Z=1$. (d) Critical behavior of the CES frequency $\omega$ close to the phase transition along the green lines in (b). The red line represents a fit to a power law. Main panel: Velocity dependence. Inset: Delta-strength dependence.}
\label{fig:PhaseDiagram}
\end{figure*}

\begin{figure*}[!tb]
\begin{tabular}{@{}ccc@{}}
    \includegraphics[width=0.66\columnwidth]{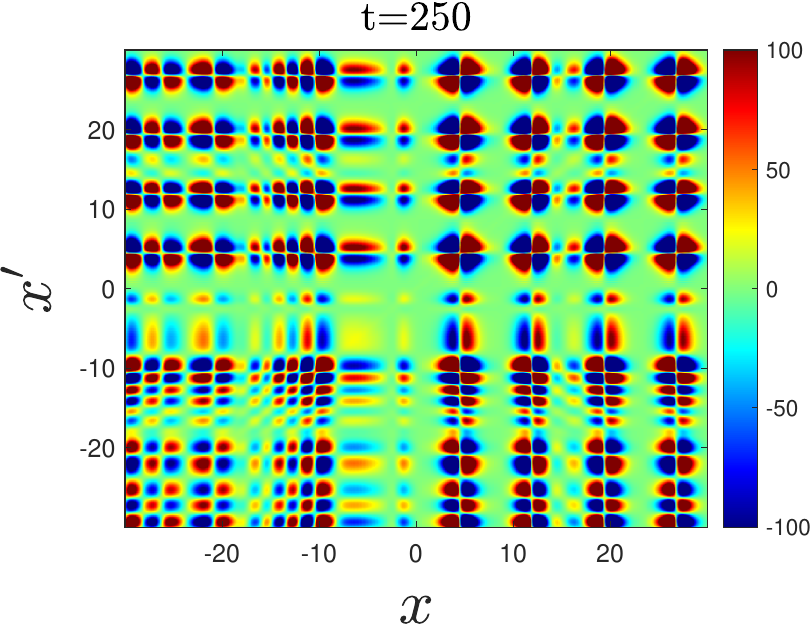} & 
    \includegraphics[width=0.66\columnwidth]{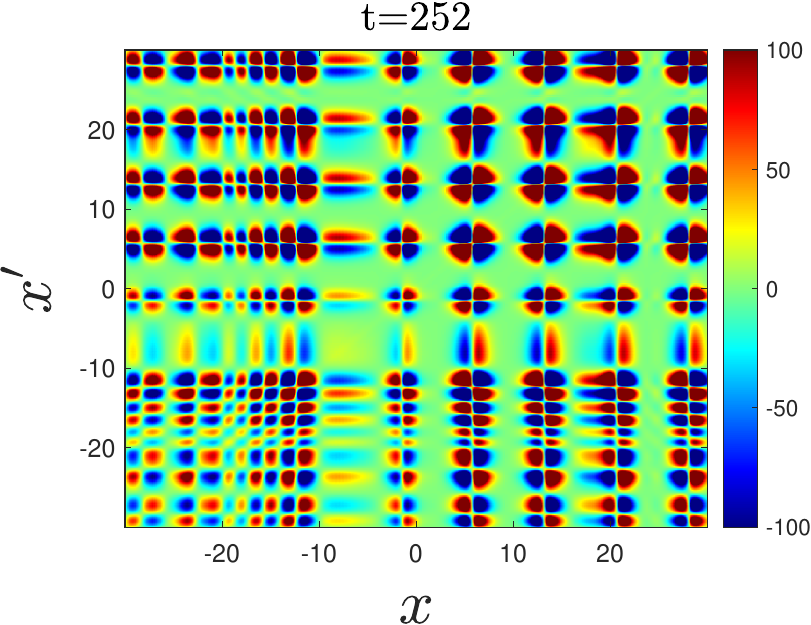} &\includegraphics[width=0.66\columnwidth]{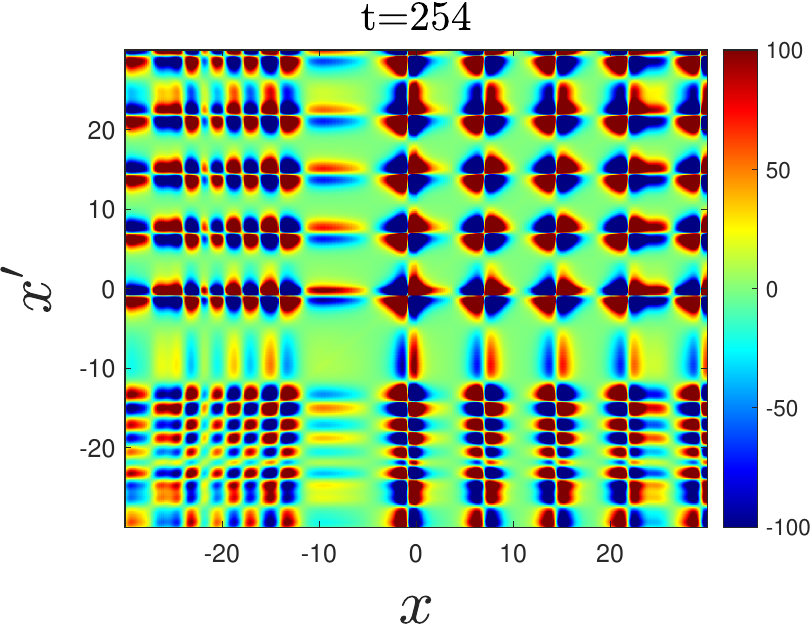}\\
   \includegraphics[width=0.66\columnwidth]{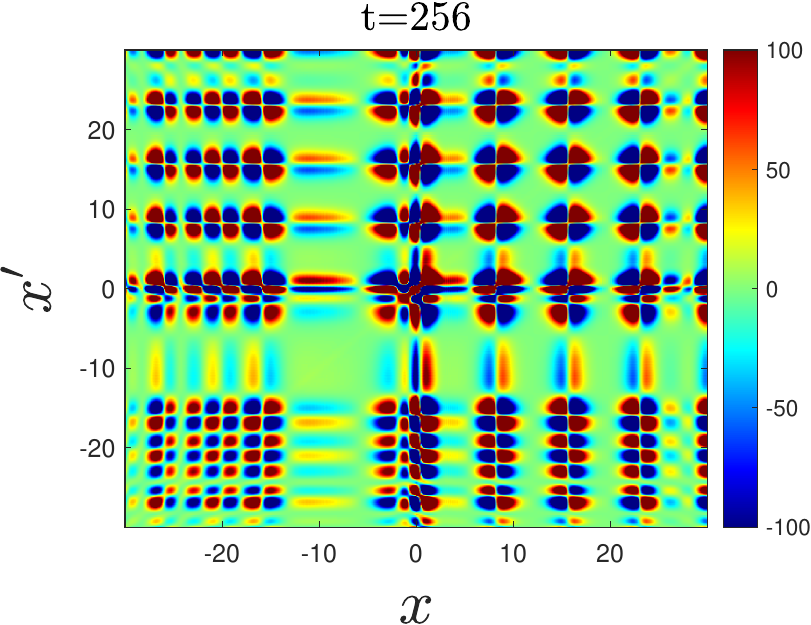} & 
    \includegraphics[width=0.66\columnwidth]{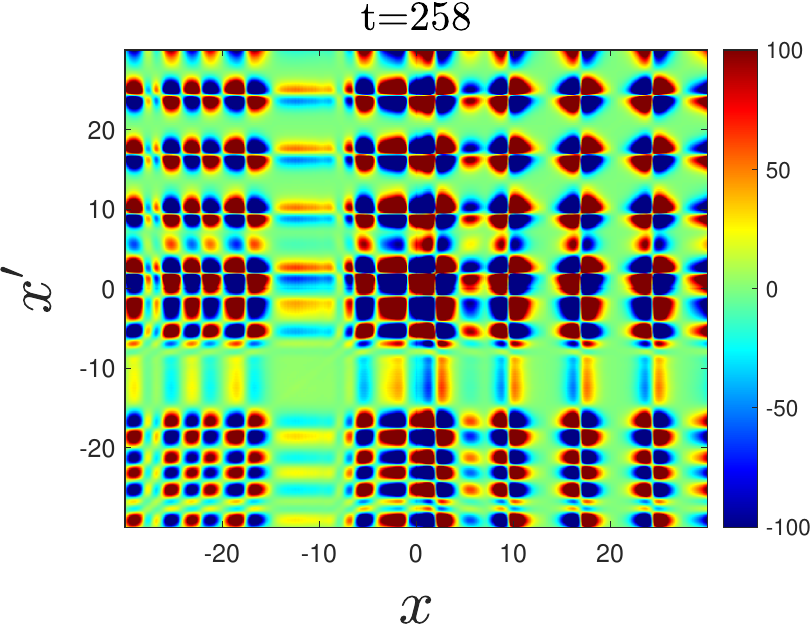} &\includegraphics[width=0.66\columnwidth]{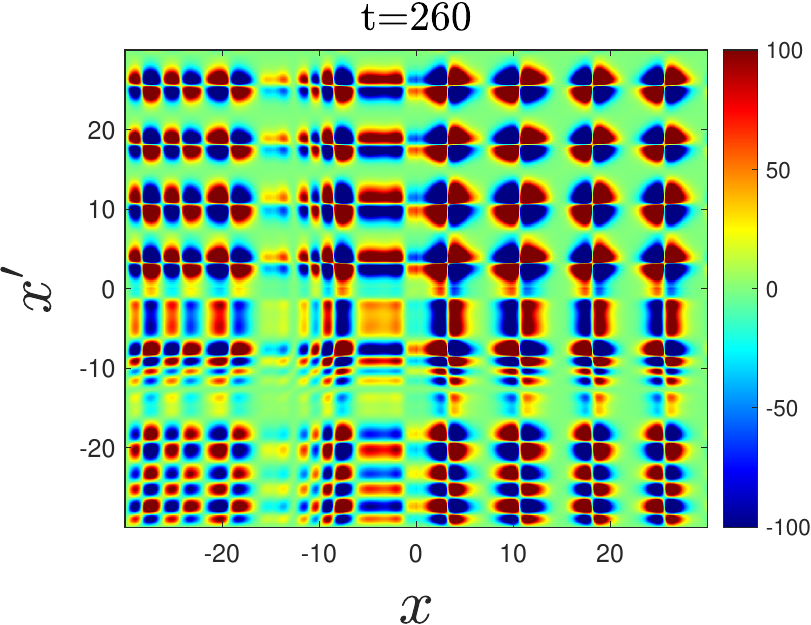}
\end{tabular}
\caption{Snapshots of the normalized density-density correlation function $G(x,x',t)$, computed using the Truncated Wigner method for a CES state with the same mean-field parameters of Fig. \ref{fig:PhaseDiagram}c, whose period is $T\approx 12.0$.}
\label{fig:QuantumCES2D}
\end{figure*}

We study in this section a specific realization of spontaneous Floquet state, the so-called CES state \cite{deNova2022}, which is a periodic solution of a 1D time-dependent GPG equation (\ref{eq:FloquetGPGTime}) of the form:
\begin{widetext}
\begin{equation}\label{eq:GPGTimeSupersolid}
    H_{GP}\Psi_0=\left[-\frac{\partial_x^2}{2}+V(x)+|\Psi_0|^2\right]\Psi_0=\mu\Psi_0+i\partial_t\Psi_0.
\end{equation} 
\end{widetext}
The presence of an inhomogeneous external potential $V(x)\neq 0$ explicitly breaks invariance under spatial translations. This guarantees the genuine character of the spontaneous symmetry-breaking of time-translation symmetry, avoiding any misidentification of time crystalline behavior as that discussed after Eq. (\ref{eq:AccidentalSFstate}). Although the CES state was originally identified in an analogue gravity context \cite{deNova2016,deNova2021}, similar scenarios involving soliton emission and periodic self-oscillations are well-known in the literature \cite{Hakim1997,Pavloff2002,Engels2007,Nguyen2017,Sels2020,Tamura2023}. 

In analogy to the case of Sec. \ref{sec:Cnoidal}, the CES state spontaneously breaks $U(1)$ and time-translation symmetries, combining superfluidity with time crystalline behavior. This represents the temporal analogue of a supersolid, which has been denoted as time supersolid and was recently observed in a magnonic BEC \cite{Autti2018}. We remark that, while strictly speaking the cnoidal wave is not a supersolid because it is not at equilibrium \cite{Martone2021}, time crystals are necessarily out-of-equilibrium states as a consequence of their celebrated no-go theorem \cite{Watanabe2015}. In the presence of external periodic driving, self-interacting condensates governed by the GP equation can also give rise to discrete time crystals \cite{Sacha2015}.

Due to the phase and time-translation invariance, if $\Psi_0(x,t)$ is a solution of Eq. (\ref{eq:GPGTimeSupersolid}), then $e^{-i\theta}\Psi_0(x,t+t_0)$ is also a solution. In the $(t,\phi)$ formalism, where $\Psi_0(x,t+t_0)=\Psi_0(x,\omega t+\phi_0)$, the simultaneous symmetry breaking in the CES state is characterized by the vectors $\boldsymbol{\alpha}=(\theta,\phi)$, $\mathbf{T}=(1,i\partial_\phi)$, $\mathbf{Q}=(N,F)$, and $\boldsymbol{\lambda}=(\mu,\omega)$. The Goldstone-Gibbs modes are then
\begin{widetext}
    \begin{align}
     \nonumber z_\theta(x,\phi)&=\left[\begin{array}{r}-i\Psi_0(x,\phi)\\ i\Psi_0^*(x,\phi)\end{array}\right],~z_\phi(x,\phi)=\left[\begin{array}{l}\partial_\phi\Psi_0(x,\phi)\\ \partial_\phi\Psi_0^*(x,\phi)\end{array}\right],\\
    \nonumber z_N(x,\phi)&=\left[\begin{array}{l}\partial_N\Psi_0(x,\phi)\\ \partial_N\Psi_0^*(x,\phi)\end{array}\right],\,z_F(x,\phi)=\left[\begin{array}{l}\partial_F\Psi_0(x,\phi)\\ \partial_F\Psi_0^*(x,\phi)\end{array}\right],\\
\end{align}
\end{widetext}
which are translated into time-dependent modes by $z_a(x,t)=z_a(x,\phi_0+\omega t)$.

The CES state satisfies typical time-crystal criteria of robustness, independence from the initial condition, and universality  \cite{deNova2022}. In this work, we focus on a specific realization in which a localized attractive delta barrier $V(x)=-Z\delta(x)$ is quenched at $t=0$ within a subsonic homogeneous condensate flowing with velocity $v$, described by a GP wavefunction $\Psi(x,0)=e^{ivx}$ (in our choice of units, the initial condensate density is then $n_0$). A schematic depiction of the initial setup is displayed in Fig. \ref{fig:PhaseDiagram}a. The quench in the external potential induces a deterministic dynamics in the condensate, numerically computed by integrating the time-dependent GP equation (\ref{eq:GPEquation}). The asymptotic behavior of the system is described by a dynamical phase diagram which is solely function of $(Z,v)$, Fig. \ref{fig:PhaseDiagram}b, exhibiting only two possible final states: the nonlinear ground state (GS) \cite{Michel2013} or the CES state. Specifically, for initial velocities larger than the critical velocity $v>v_c(Z)$, or, equivalently, for barrier amplitudes larger than the critical amplitude $Z>Z_c(v)$, the GP  wavefunction asymptotically approaches the CES state along the lines of Eq. (\ref{eq:GP2GPGTime}),
\begin{equation}
    \Psi(x,t)\xrightarrow[t\to\infty]{} e^{-i\mu t}\Psi_0(x,t),
\end{equation}
where it oscillates periodically. This periodic behavior is globally displayed both upstream and downstream, as seen in Fig. \ref{fig:PhaseDiagram}c, and not just in the traveling cnoidal wave downstream (whose time periodicity is rather trivial, as previously argued). We note that the CES state is a macroscopically excited state, something expected from the discussion at the end of Sec. \ref{subsec:TimeOperator}, since the initial state is shifted by an energy $\Delta E=Nv^2/2$ with respect to a condensate at rest. 

The GS/CES phase diagram is an example of dynamical phase transition \cite{Moeckel2008,Sciolla2010,Lang2018}, where the GS is the symmetry-unbroken phase, with continuous time-translation symmetry, while the CES state is the time-crystalline phase, with discrete time-translation symmetry. The oscillation frequency $\omega$ exhibits a critical behavior close to the phase boundary, where the critical exponents $\delta_v,\delta_Z$ for $v,Z$ (obtained from a fit in Fig. \ref{fig:PhaseDiagram}d) are both approximately $\delta_v\simeq \delta_Z\simeq 0.50$, in agreement with those for a square well \cite{deNova2022}, potentially suggesting a possible analytical derivation. The main advantage of using a delta barrier with respect to a finite square well is that the transient towards the CES state is shorter, reducing the numerical cost of the simulations.

\begin{figure*}[!tb]
\begin{tabular}{@{}cc@{}}
     \stackinset{l}{0pt}{t}{0pt}{\textbf{(a)}}{\includegraphics[width=\columnwidth]{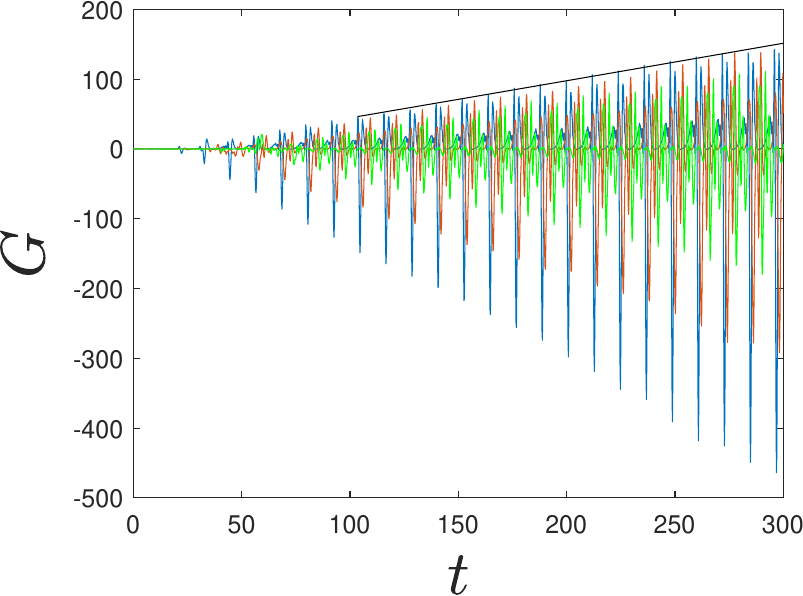}} ~~&~~ 
     \stackinset{l}{0pt}{t}{0pt}{\textbf{(b)}}{\includegraphics[width=\columnwidth]{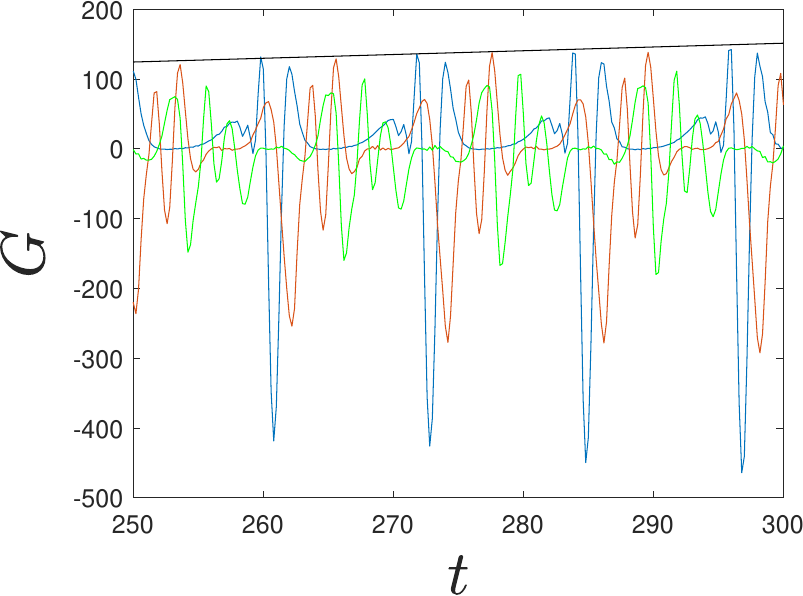}} \\ \\ \\
     \stackinset{l}{0pt}{t}{0pt}{\textbf{(c)}}{\includegraphics[width=\columnwidth]{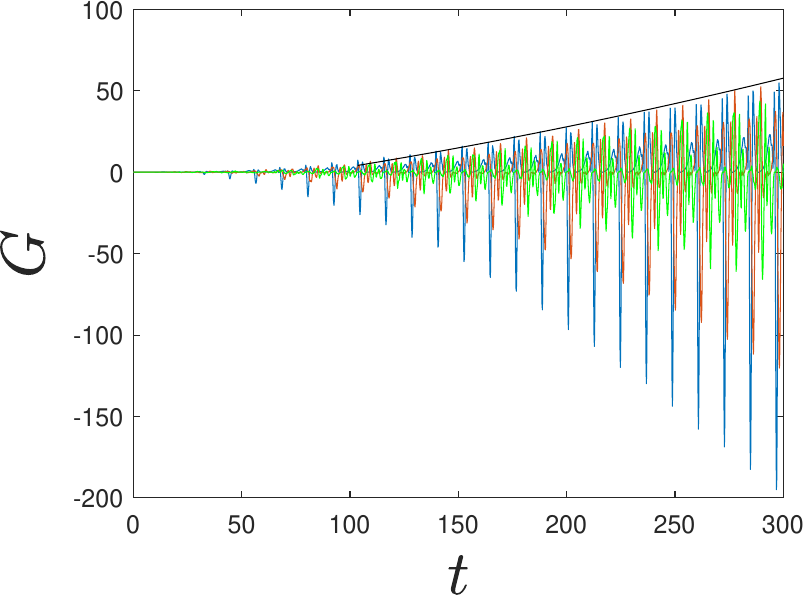}} ~~&~~ 
     \stackinset{l}{0pt}{t}{0pt}{\textbf{(d)}}{\includegraphics[width=\columnwidth]{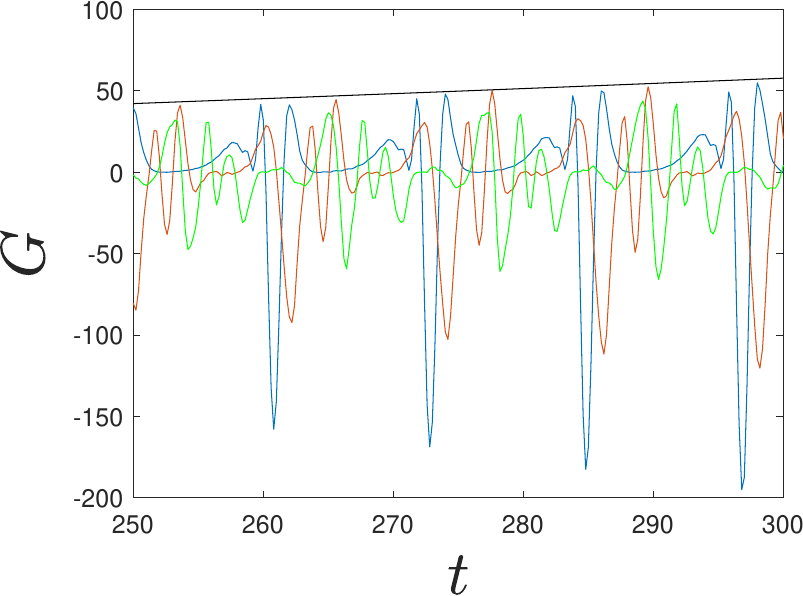}}
\end{tabular}
\caption{(a) Normalized density-density correlation function $G(x,-x,t)$ for $x=10$ (blue), $x=20$ (red), and $x=30$ (green) for the simulation in Fig. \ref{fig:QuantumCES2D}.  Black line is a quadratic polynomial fit to the amplitude of the oscillations, $A(t)=at^2+bt+c $. (b) Zoom of (a). (c)-(d) Same as upper row but now replacing quantum fluctuations by particle-number fluctuations. Specifically, we take as initial condition in the Truncated Wigner method $\Psi(x,0)=\sqrt{1+\delta n}\,e^{ivx}$, with $\delta n$ a Gaussian random variable with zero mean and $\sqrt{\braket{\delta n^2}}=0.001$.}
\label{fig:QuantumParticleCompa1D}
\end{figure*}



We now go beyond mean-field and explicitly compute the dynamics of the quantum fluctuations by using the Truncated Wigner method \cite{Sinatra2002,Carusotto2008}; all technical details behind the numerical simulations and the specific implementation of the Truncated Wigner method can be found in Refs. \cite{deNova2016,deNova2022,deNova2023}. As initial quantum state, we take the $T=0$ ground state in the comoving frame of the condensate. Regarding the observables of interest, we focus on computing the \textit{normalized} density-density correlation function 
\begin{equation}
    G(x,x',t)\equiv \frac{\braket{\delta\hat{n}(x,t)\delta\hat{n}(x',t)}}{n_0\xi^{-1}_0},
\end{equation}
with $\delta\hat{n}(x,t)\equiv \hat{n}(x,t)-n(x,t)$ the fluctuations around the ensemble-averaged density $n(x,t)\equiv\braket{\hat{n}(x,t)}$, and $\xi_0$ the healing length associated to the initial density $n_0$; we have momentarily restored dimensions in this expression for the sake of clarity. In the laboratory, density-density correlations can be measured through \textit{in situ} imaging after averaging over ensembles of repetitions of the experiment \cite{Steinhauer2016,deNova2019,Kolobov2021}.

After some transient, and once in the CES state, the correlations exhibit a seemingly periodic behavior, as shown in Fig. \ref{fig:QuantumCES2D}, where sharp features arise due to the synchronized emission of shock waves/solitons into the upstream/downstream region. Further insight is obtained in upper row of Fig. \ref{fig:QuantumParticleCompa1D}, where we plot the time dependence of the correlation function $G(x,-x,t)$ between symmetric upstream and downstream points for different values of $x$. We observe that the correlations are indeed quasi-periodic, displaying periodic oscillations whose amplitude grows in time. 

\begin{figure*}[t]
\begin{tabular}{@{}ccc@{}}
    \includegraphics[width=0.66\columnwidth]{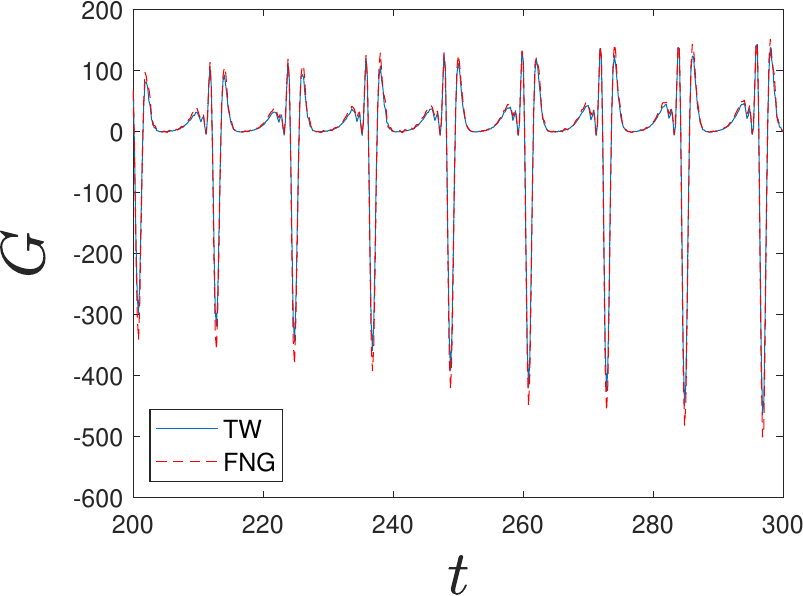} & \includegraphics[width=0.66\columnwidth]{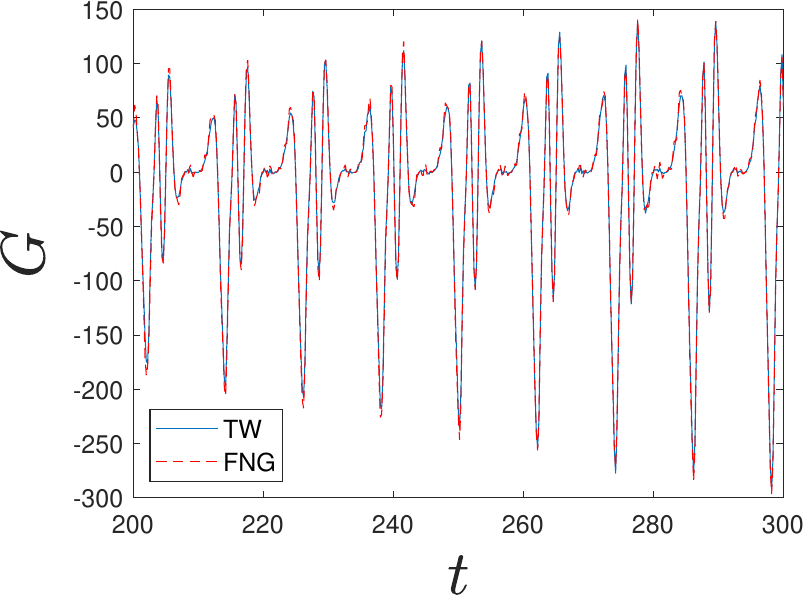} & \includegraphics[width=0.66\columnwidth]{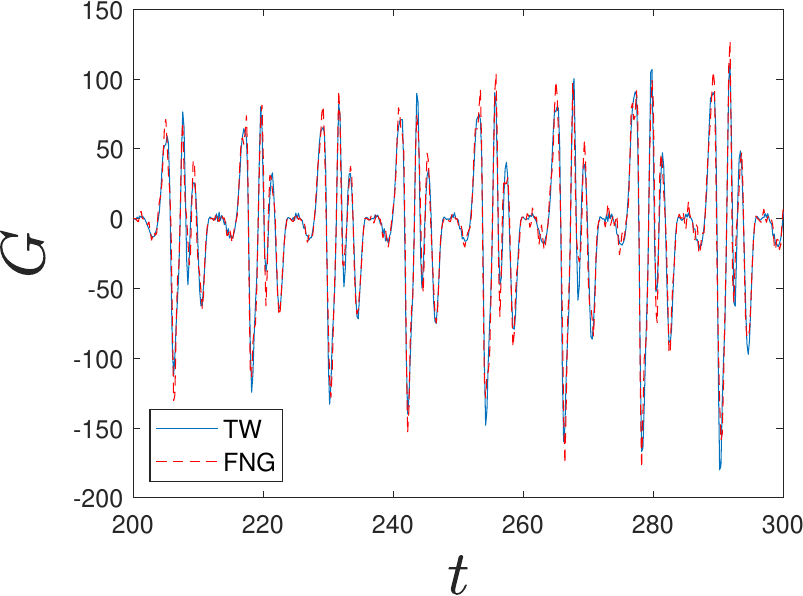}\\ \\ 
    \includegraphics[width=0.66\columnwidth]{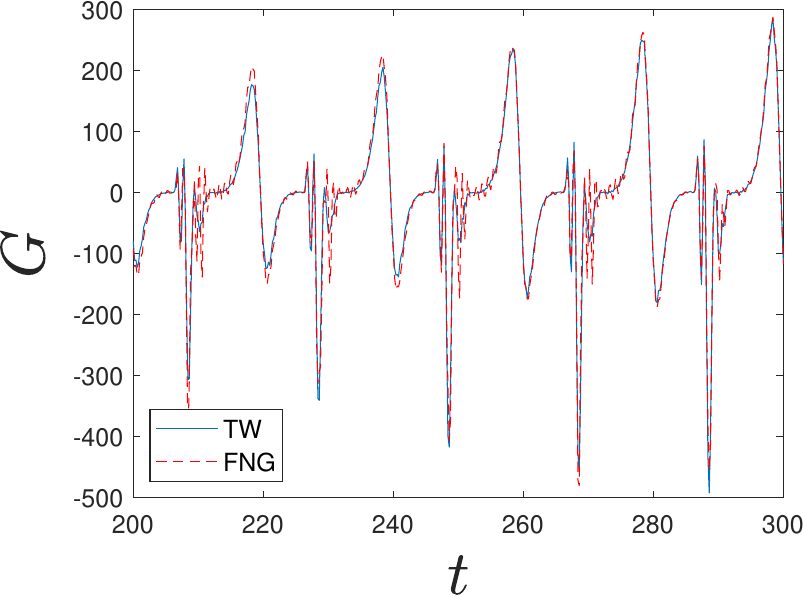} & \includegraphics[width=0.66\columnwidth]{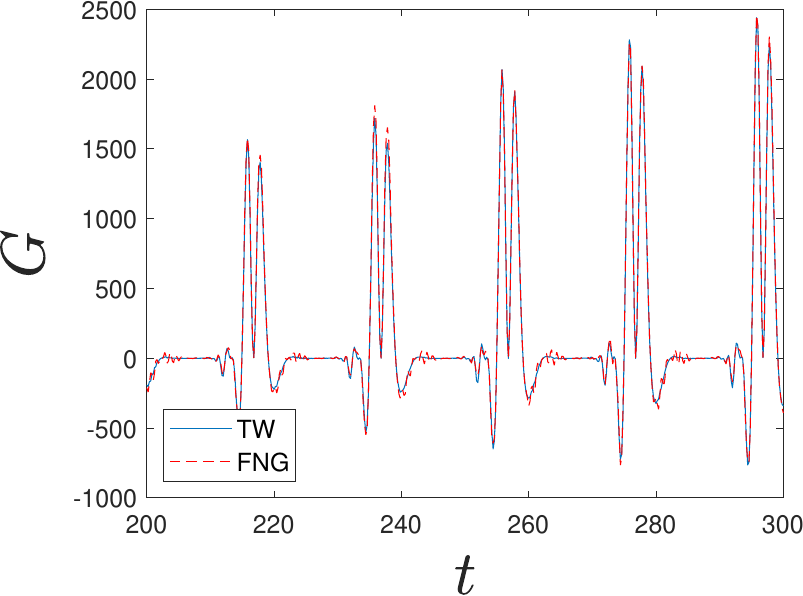} & \includegraphics[width=0.66\columnwidth]{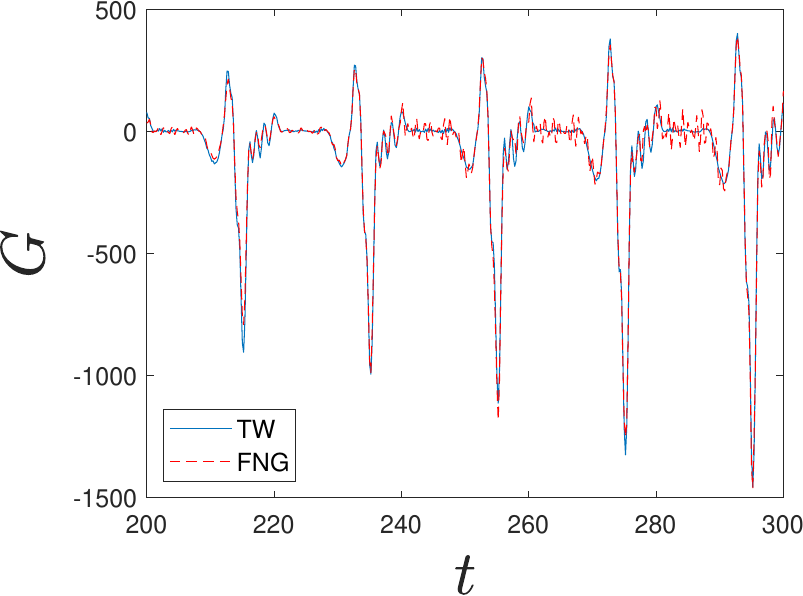}
\end{tabular}
\caption{Normalized density-density correlation function $G(x,-x,t)$ for $x=10$ (left column), $x=20$ (center column), and $x=30$ (right column). Solid blue is the numerical result from a Truncated Wigner simulation and dashed red is the contribution from the temporal FNG mode,  $G(x,x',t)=A(t)r_t(x,t)r_t(x',t)$, where $A(t)$ is numerically fitted as in Fig. \ref{fig:QuantumParticleCompa1D} and $r_t(x,t)=\partial_t n_0(x,t)$ is computed from the time derivative of the mean-field CES density. Upper row: $Z=1$ and $v=0.8$. Lower row: $Z=1$ and $v=0.7$.}
\label{fig:TWFG}
\end{figure*}

\begin{widetext}
This behavior can be understood using the results of Sec. \ref{sec:SimultaneousSymmetryBreakingFloquet}. After expanding the field operator around the CES wavefunction, we find that the density fluctuations read at linear order in the field fluctuations $\hat{\varphi}(x,t)$ as
\begin{equation}
    \delta\hat{n}(x,t)\simeq\Psi_0^*(x,t)\hat{\varphi}(x,t)+\Psi_0(x,t)\hat{\varphi}^\dagger(x,t)=-iz^\dagger_\theta \sigma_z \hat{\Phi}.
\end{equation}
Plugging the expansion of Eq. 
(\ref{eq:FloquetFieldPhiFluctuationsTimeDependent}) into this expression yields
\begin{equation}\label{eq:DensityFieldFluctuations}
   \delta\hat{n}(x,t)=\hat{X}^t(t) r_t(x,t)+ \hat{P}^Nr_N(x,t)+\hat{P}^Fr_F(x,t)\\
   +\sum_{n}\hat{\gamma}_{n}r_{n}(x,t)e^{-i\varepsilon_nt}+\hat{\gamma}^{\dagger}_{n}r^*_{n}(x,t)e^{i\varepsilon_nt},
\end{equation}  
\end{widetext}
$r(x,t)=-iz^\dagger_\theta(x,t)\sigma_z z(x,t)$ being the density wavefunction associated to the spinor $z$. In the case of the Goldstone-Gibbs modes, the density wavefunction is simply obtained by 
\begin{equation}
r_a(x,\phi)=\partial_a|\Psi_0(x,\phi)|^2\equiv\partial_a n_0(x,\phi),
\end{equation}
where $n_0(x,\phi)$ is the mean-field CES density, which is approximately equal to the ensemble-averaged density after neglecting higher-order corrections in the field fluctuations, $n(x,t)\simeq n_0(x,t)=n_0(x,\phi_0+\omega t)$.

As physically expected, the FNG mode associated to the phase does not couple to the density, $r_\theta=0$. Thus, only the temporal FNG mode $z_t(x,t)$ contributes to the correlation function $G(x,x',t)$. Moreover, its amplitude $\hat{X}^t(t)=\hat{X}^\phi(t)/\omega$ grows linearly in time according to Eq. (\ref{eq:TimeOperatorEvolution}), so it eventually dominates the correlations,
\begin{align}\label{eq:PredictionCorrelationFloquet}
      G(x,x',t)&\simeq  A(t)r_t(x,t)r_t(x',t),\\
     \nonumber A(t)&\equiv\frac{\braket{(\hat{X}^t(t))^2}}{n_0\xi^{-1}_0}.
\end{align}
This is a quasi-periodic time dependence, where $r_t(x,t)r_t(x',t)$ is a periodic term as $r_t(x,t)=\partial_t n_0(x,t)$, while the global amplitude $A(t)$ encodes the non-periodic part. Specifically, due to the ballistic evolution of $\hat{X}^t(t)$, we expect a quadratic polynomial dependence, $A(t)=at^2+bt+c$. This fits well the observed growth, black lines in upper row of Fig. \ref{fig:QuantumParticleCompa1D}. A more detailed comparison is presented in Fig. \ref{fig:TWFG}, where we depict together the numerical results for $G(x,x',t)$ from the Truncated Wigner method (solid blue), and the theoretical prediction from Eq. (\ref{eq:PredictionCorrelationFloquet}), combining the previous quadratic fit of $A(t)$ with a computation of $r_t(x,t)$ from the time derivative of the mean-field density (dashed red). A remarkable agreement is observed, confirming the ballistic motion of the quantum amplitude of the temporal FNG mode as the mechanism responsible for the quasi-periodic correlations.

We note that the coefficients $a,b,c$ determining the amplitude $A(t)$ are given in terms of expectation values which are quadratic in $\hat{X}^t,\hat{P}^N,\hat{P}^F$, evaluated in the initial quantum state. Thus, our specific time-dependent scheme eliminates any possible ambiguity in the definition of the quantum state of the system \cite{Ribeiro2022}. Nevertheless, fluctuations of the temporal FNG can be also induced by run-to-run experimental variations such as shot noise in the initial particle number. This is shown in lower row of Fig. \ref{fig:QuantumParticleCompa1D}, where we display the results of solely including fluctuations in the particle number of the initial condition. The observed correlation patterns are the same as in the purely quantum case (upper row), further revealing their monomode origin since particle-number fluctuations are structureless \cite{deNova2023}.

\begin{figure}[t]
    \includegraphics[width=\columnwidth]{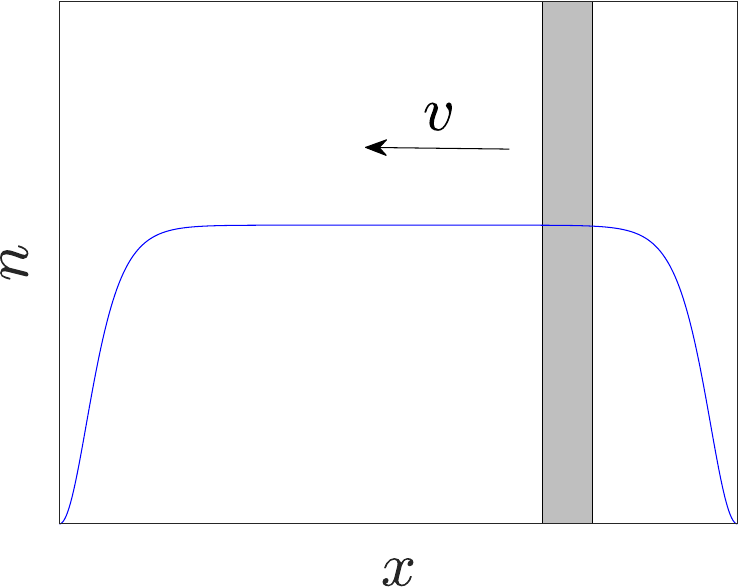} 
\caption{Schematic representation of the proposed experimental setup to observe the FNG mode. A localized obstacle (indicated by the shaded area) is swept with velocity $v$ along the homogeneous bulk of a condensate confined by a boxlike potential (whose initial density profile is given by solid blue line). By Galilean invariance, this is equivalent to launching the condensate over the obstacle with the same velocity. When $v$ is above the critical velocity, the CES state is achieved at long times; then its density correlations can be characterized through high-resolution imaging.} 
\label{fig:Expeme}
\end{figure}

Based on these results, we propose an experimental configuration to observe the temporal FNG mode involving standard techniques in analogue gravity setups \cite{Steinhauer2016,deNova2019,Kolobov2021}, schematically depicted in Fig. \ref{fig:Expeme}. We consider an elongated quasi-1D condensate confined by a boxlike potential \cite{Meyrath2005,Van2010,Gaunt2013,Rauer2018,Murtadho2025}, whose density is essentially homogeneous far from the edges. At $t=0$, a localized obstacle (indicated by the shaded area) is swept with velocity $v$ along the bulk of the condensate. By Galilean invariance, this is equivalent to launching the condensate against the obstacle with the same velocity, achieving a configuration similar to that in Fig. \ref{fig:PhaseDiagram}a. For values of $v$ above the critical velocity, a CES state will be reached at long times. Another possibility is to confine the condensate in a long ring \cite{Eckel2018} and rotate a localized potential. In both cases, at late times once in the CES regime, when the temporal FNG mode dominates the Bogoliubov dynamics, high-resolution imaging can be then used to measure the time evolution of the density profile $n(x,t)$ and the density-density correlations $G(x,x',t)$, from where the FNG mode can be extracted as in Figs. \ref{fig:QuantumCES2D}-\ref{fig:TWFG}.


We note that the protocol used here to reach the CES state is fully deterministic at the mean-field level, which imposes the value of the global time origin $t_0$ of the spontaneous Floquet state. Thus, even though there is a temporal FNG mode, technically speaking the symmetry breaking of time-translation invariance is not spontaneous. The reason is that our scheme fixes a $t=0$ origin in the GP equation. Regardless, at late times, due to interactions, the system enters the CES self-oscillating regime where it forgets about the initial condition and exhibits all the features of a spontaneous Floquet state, including the temporal FNG mode. We can qualitatively understand this behavior by drawing an analogy with a conventional spatial crystal growing in the $z$-direction on top of a rigid substrate, which fixes a $z=0$ plane that breaks continuous spatial translation symmetry. In the bulk of the grown crystal, however, the system becomes insensitive to the boundary and recovers the usual properties of a crystalline structure. 

Nonetheless, the phase locking of $\phi_0$ actually helps to measure the FNG mode since, if $\phi_0$ were spontaneously chosen in each realization of the experiment, then the ensemble-averaged value of $G(x,x',t)$ would represent an angular average over $\phi_0\in[0,2\pi)$, therefore suppressing its time dependence. In that case, one would need to resort to the complete \textit{out-of-time} correlation function
\begin{equation}
    G(x,x',t,t')\equiv \frac{\braket{\delta\hat{n}(x,t)\delta\hat{n}(x',t')}}{n_0\xi^{-1}_0},
\end{equation}
which requires the use of non-destructive imaging in order to measure the density at two consecutive times $t>t'$, as in Ref. \cite{Smits2018}. This caveat is not needed for the measurement of correlations under the usual $U(1)$ symmetry breaking, where each realization of the experiment is insensitive to the global condensate phase because observables always commute with the total particle number, the generator of global phase transformations.

\section{Discussion and perspectives}\label{sec:Discussion}

We proceed to critically discuss the results of this work and how they inscribe within the literature, outlining in the process future avenues. We have examined in Sec.~\ref{sec:SimultaneousSymmetryBreaking} how the general results from Goldstone theorem applied to variational descriptions, Sec.~\ref{sec:GoldstoneGeneral}, can be translated to many-body systems when several symmetries are simultaneously broken by the quantum state, focusing on the specific case of a BEC close to zero temperature for illustrative purposes. We have found that a rather complete description of the problem is provided in terms of the generalized Gibbs ensemble, which treats all broken symmetries on equal grounds. The Lagrange multipliers associated to each conserved charge behave as generalized velocities along the orbits of the symmetry transformations, extending the well-known result that the chemical potential is the velocity of the phase. We have also proven fundamental results in Thermodynamics for the symmetry-broken state, such as the first principle or the Gibbs-Duhem relation.

The Nambu-Goldstone modes associated to each spontaneously broken symmetry emerge as zero-energy modes of the excitation spectrum. Each NG mode is paired with a Gibbs mode, describing the fluctuations of the quantum state with respect to the associated conserved charge. These fluctuations arise because a symmetry-broken state cannot be eigenstate of the corresponding charge operators. The orthogonalization of the Goldstone-Gibbs modes involves the Berry-Gibbs connection, a Berry connection whose variables are the conserved charges, which are not intrinsic parameters of the Hamiltonian. In contrast to the standard case, the resulting Berry-Gibbs curvature is not invariant under generalized gauge transformations, expanding the usual notion of phase transformations to include more general symmetry transformations. When extended to the complete manifold that includes the continuous parameters of the spontaneously broken symmetries, the Berry-Gibbs curvature gives rise to a symplectic structure that allows to quantize the amplitudes of the Nambu-Goldstone/Gibbs modes as conjugate coordinate/momentum operators. This analogy is translated to the dynamics, where the momenta are time-independent as they represent the quantum fluctuations of the corresponding conserved charges, while the coordinates display a ballistic motion with velocities that depend linearly on the momenta. The physical intuition behind this dynamics is very simple: since there is no restoring force along the orbits generated by the broken-symmetry transformations, fluctuations in those directions, spanned by the NG modes, lead to unbounded motion with constant velocity. Finally, we show that the Goldstone-Gibbs sector can be diagonalized, leading to a new set of independent NG modes. Remarkably, this diagonalization only involves thermodynamic considerations, and describes how the original NG modes hybridize within the different branches of the spectrum in the low-energy limit.

The coordinate-momentum analogy between the Nambu-Goldstone and Gibbs amplitudes was originally established in Ref. \cite{Lewenstein1996} in the context of the $U(1)$-symmetry breaking and the phase diffusion of a condensate. This approach was followed in Ref. \cite{Dziarmaga2004} to also quantize the position of a soliton and study its quantum diffusion, inspiring a whole line of research on quantum solitons \cite{Sacha2009,Gangardt2010,WadkinSnaith2012,Walczak2011,Marchukov2020}. A study of the quantization of arbitrary NG modes in condensates, including internal symmetries such as spin, was provided in Ref. \cite{Takahashi2015}. The present work further extends those results by developing a general and systematic framework with the help of the generalized Gibbs ensemble and the Berry-Gibbs connection, which in addition allows to understand the physics of the Nambu-Goldstone and Gibbs modes at a more fundamental level. On the other hand, the emergence of a geometrical description and a symplectic structure in variational formulations is well-known in the literature and can be traced back to Ref. \cite{Kramer2005} (see Ref. \cite{Kramer2008} for a more recent review), where the manifold is generated there by the variational parameters of the ansatz and their velocities, leading to a \textit{dequantization} of the problem. In contrast, the manifold variables in our description are the continuous parameters of the spontaneously broken symmetries and their associated charges, instead of variational parameters, and the symplectic structure conversely leads to a \textit{quantization} of the Goldstone-Gibbs amplitudes. To the best of our knowledge, the thermodynamical and geometrical framework developed in Sec. \ref{sec:SimultaneousSymmetryBreaking} for the description of spontaneous symmetry breaking and the quantization of the Goldstone-Gibbs modes is original from this work and is not present in the literature. 

As a particular application of this formalism, we have studied in Sec.~\ref{sec:Cnoidal} the case of a cnoidal wave in a superfluid, which spontaneously breaks $U(1)$ and spatial translation symmetry, extending the work of Refs. \cite{Dziarmaga2004,Martone2021} into the generalized Gibbs ensemble. We have explicitly checked the consistency and validity of the theoretical predictions of Sec.~\ref{sec:SimultaneousSymmetryBreaking}. 

The results of Secs.~\ref{sec:SimultaneousSymmetryBreaking},\,\ref{sec:Cnoidal} are thus of interest for the study of states which simultaneously break several continuous symmetries, such as supersolids, quantum droplets, spinor and rotating condensates, or quantum solitons. In particular, they find immediate application in systems described by extended nonlinear equations beyond the usual Gross-Pitaevskii description, such as the Lee-Huang-Yang correction  \cite{Petrov2015,Cheiney2018,Ancilotto2018} or non-polynomial Gross-Pitaevskii equations \cite{Salasnich2002,deNova2019}. Our results can also be easily translated to magnonic condensates \cite{Makinen2024}, nonlinear optical fibers \cite{Drummond2014}, and quantum fluids of light \cite{Carusotto2013}, as these obey similar nonlinear equations of motion. Indeed, the generality of the formalism, based on a Lagrangian description, allows its adaptation to arbitrary many-body systems, including fermions. For instance, within the $\nu=0$ quantum Hall phase diagram in graphene, states with two spontaneously broken symmetries were predicted within a Hartree-Fock description \cite{deNova2017}, of which some experimental evidence was recently reported \cite{Coissard2022}. Indeed, Ref. \cite{deNova2017} established the correspondence between the BdG and TDHFA equations, and translated Goldstone theorem into the specific framework of the Hartree-Fock equations. Therefore, the BdG/TDHFA correspondence provides a direct pipeline through which our results can be exported. 

Another interesting perspective is provided by the geometry of the extended Berry-Gibbs connection. In this work, we have assumed that all the generators of the spontaneously broken symmetries commute between themselves. This leads, after a generalized gauge transformation, to a ``flat'' symplectic curvature $F_{ab}=-\Omega_{ab}$. Actually, a generalized gauge transformation is nothing else but a change of coordinates in the extended Berry-Gibbs manifold. However, if the generators of the spontaneously broken symmetries form a non-abelian Lie algebra, this no longer holds, opening the door for deeper geometrical implications, including non-trivial topological aspects. The exploration of this promising avenue is left for future work. 

The formalism developed for stationary states in Secs.~\ref{sec:GoldstoneGeneral},\,\ref{sec:SimultaneousSymmetryBreaking} is adapted to Floquet states in Secs.~\ref{sec:GoldstoneFloquetGeneral},\,\ref{sec:SimultaneousSymmetryBreakingFloquet}. Specifically, from the general results of Sec.~\ref{sec:GoldstoneGeneral}, we prove that the usual Nambu-Goldstone modes with zero energy are now translated into Floquet-Nambu-Goldstone modes with zero quasi-energy. Moreover, since spontaneous Floquet states break continuous time-translation symmetry, they possess a genuine temporal FNG mode. This is a distinctive signature of a spontaneous Floquet state, completely absent in conventional Floquet systems.

In order to correctly describe the dynamics of the FNG modes of a spontaneous Floquet state, we develop the $(t,\phi)$ formalism, which goes beyond the usual $(t,t')$ formalism. In combination with the generalized Gibbs ensemble, we provide a thermodynamical description of Floquet states analogous to that of stationary states, which we refer to as Floquet thermodynamics. This is possible because spontaneous Floquet states also have a well-defined energy, which enables the existence of a conserved Floquet charge. The use of the phase $\phi$ instead of the periodic time $t'$ is essential to both identify the thermodynamical role of the Floquet charge $Q_\phi=F$ and the dynamical role of its Lagrange multiplier $\lambda^\phi=\omega$. The Floquet charge is another hallmark of a spontaneous Floquet state, which adds to the presence of a temporal FNG mode. In fact, both features are the sides of the same coin, since they arise, via Noether and Goldstone theorems, from the continuous time-translation symmetry of the underlying time-independent Hamiltonian. 

Conventional Floquet states can also be described within the framework of Floquet thermodynamics by means of the Floquet enthalpy, i.e., the Legendre transform of the energy with respect to the Floquet charge, whose conjugated variable is in turn the frequency, fixed here by the external driving. By drawing an analogy with standard Thermodynamics, spontaneous Floquet states are ``isofloquetic'', and driven Floquet states are ``isoperiodic''.

The quantization procedure of the Goldstone-Gibbs sector goes along the same lines as in the stationary case, involving a Berry-Gibbs connection and a symplectic form. Their dynamics also follow the coordinate/momentum picture. Nevertheless, we now have a novel temporal FNG mode, whose amplitude can be regarded as an effective time coordinate, with a conjugate momentum given by the energy fluctuations. Therefore, we can identify the quantum amplitude of the temporal FNG mode as a unique realization of a time operator in Quantum Mechanics.


Finally, we apply our formalism to a specific realization of spontaneous Floquet state, the CES state, which spontaneously breaks $U(1)$ and continuous time-translation symmetry, representing a time supersolid. We propose an experimental scheme, based on standard analogue gravity techniques \cite{Steinhauer2016,deNova2019,Kolobov2021}, to observe the temporal FNG mode, which does couple to the density in contrast to the $U(1)$ FNG mode. In particular, we prove that the density-density correlations are dominated at long times by the ballistic growth of the temporal FNG mode, from where it can be extracted. Numerical results from a Truncated Wigner simulation show a remarkable agreement with our theory.

The results of Secs.~\ref{sec:GoldstoneFloquetGeneral},\,\ref{sec:SimultaneousSymmetryBreakingFloquet} can be used to describe simultaneous symmetry breaking in both spontaneous and conventional Floquet states. Furthermore, the $(t,\phi)$ formalism and the Floquet thermodynamics can be applied to an arbitrary many-body system, where they establish a fundamental distinction between both states at the thermodynamical level. In the specific case of the CES state, this survives for thermodynamically long times which scale linearly with the system size \cite{deNova2022}. This can be compared with the prethermal regime of a driven Floquet system, whose lifetime scales exponentially with the applied frequency \cite{Pizzi2021,Ye2021,Kyprianidis2021}. From a broader perspective, the description of general quantum systems in terms of traditional thermodynamic tools is an active topic of research \cite{Allahverdyan2004,Bera2019,Monsel2020,Elouard2023}.


The quest for a time operator is a fundamental problem in Quantum Mechanics \cite{Aharonov1961,Susskind1964,Unruh1989,Muga2007}, which is still open \cite{Honh2021}. In our case, the time operator emerges from the quantum fluctuations of the global time origin, in close analogy with the fluctuations of the global phase of a condensate. Indeed, in the $(t,\phi)$ formalism, the time operator emerges from the fluctuations of the global phase shift $\phi_0$ in the periodic motion of the spontaneous Floquet state. The connection between time and phase has been pointed out since early approaches to the subject \cite{Susskind1964} because any realization of a time operator is expected to be based on the oscillation of a clock. The time-phase analogy suggests that spontaneous Floquet states should be then macroscopically excited states, with an energy well above the ground state. As a result, the CES state provides a condensed-matter setup where a tangible realization of a time operator can be studied. In general, each specific proposal of time operator presents its own characteristic features, determined by the underlying system acting as a clock, such as those based on kaons \cite{Bramon2004,Durt2019,Bernabeu2022} or the so-called time of arrival
\cite{Kijowski1974,Grot1996,Delgado1997,Aharonov1998}. 

Our work also finds a major application in the field of time crystals, since a spontaneous Floquet state represents a specific realization of continuous time crystal \cite{deNova2022}. Indeed, some results of our work are connected with the time-crystal literature. For instance, the role of the generalized Gibbs ensemble in dissipative time crystals was thoroughly discussed in Ref. \cite{Booker2020}. The emergence of continuous time crystals from excited eigenstates, in a similar fashion to spontaneous Floquet states, was predicted in Ref. \cite{Syrwid2017}. Time supersolids, like the CES state, were originally observed in Ref. \cite{Autti2018}, to which the formalism of this work is directly applicable. Our explicit characterization of the temporal FNG mode complements other field-theoretical examples present in the literature \cite{Hayata2018,Hongo2021,Daviet2024,Daviet2025}. On the other hand, Floquet thermodynamics can provide a novel tool for the characterization of both continuous and discrete time crystals. Moreover, the analysis presented in Sec.~\ref{sec:Cnoidal} suggests that certain periodic systems identified as time crystals may not genuinely break continuous time-translation symmetry since they can arise from constant motion along a closed orbit generated by other broken-symmetry transformations. As a result, there is not a truly temporal NG mode since this can be expressed in terms of the remaining NG modes. Thus, the presence of a genuine temporal FNG mode should be regarded as the characteristic signature of any \textit{bona fide} continuous time crystal. 

Extended Gross-Pitaevskii equations are also expected to exhibit spontaneous Floquet states. Although we have used atomic condensates to illustrate our Floquet formalism, this can be straightforwardly extrapolated to magnonic condensates, nonlinear optical fibers or quantum fluids of light. Spontaneous Floquet states in fermionic systems described by Hartree-Fock approaches, such as quantum Hall states, are also plausible \cite{deNova2022}. In addition, self-consistent BCS-type theories are known to display periodic oscillations in momentum space \cite{Barankov2004,Foster2014,Perfetto2020}. Spontaneous Floquet states can also arise in few-body \cite{Silveirinha2023} and many-body \cite{Dreon2022} dissipative quantum systems whose effective dynamics is described by
self-consistent equations of motion. In general, spontaneous Floquet states extend the field of nonlinear Floquet waves \cite{Kreil2019,Trager2021} to scenarios without external driving. 




Future extensions of the present work should include finite-temperature effects and quantum corrections beyond linearity, both for stationary and Floquet states exhibiting simultaneous symmetry breaking. Of particular interest is the general study of the nonperturbative quantum diffusion of the spontaneously broken symmetries, generalizing the works of Refs. \cite{Lewenstein1996,Dziarmaga2004}. In the case of the temporal Floquet-Nambu-Goldstone mode, this could lead to suggestive concepts such as time diffusion or time eigenstates, in analogy with the phase  \cite{Leggett1991,Sols1994a}. A fully quantum time crystal based on a spontaneous Floquet state, spontaneously breaking time-translation symmetry without phase-locking its time origin as in the present work, can be achieved through the self-amplification of quantum Hawking radiation \cite{deNova2025}.


\acknowledgments

We thank C. Cabrera, S. Erne, M. A. Garc\'ia-March, A. Haller, N. Hosseini, C.-L. Hung, and Y. Zenati for useful discussions. This project has received funding from European Union's Horizon 2020 research and innovation programme under the Marie Sk\l{}odowska-Curie Grant Agreement No. 847635, from Spain's Agencia Estatal de Investigaci\'on through Grant No. PID2022-139288NB-I00, and from Universidad Complutense de Madrid through Grant No. FEI-EU-19-12.

\appendix

\section{Cnoidal waves in the GPG equation}\label{app:GPGCnoidals}

\subsection{Elliptic functions}

We briefly review here some basic notions of elliptic functions. We begin by defining the Jacobi elliptic functions
\begin{eqnarray}
    \textrm{sn}(u,\nu)&\equiv&\sin\left(\textrm{am}(u,\nu)\right),\\
    \nonumber \textrm{cn}(u,\nu)&\equiv&\cos\left(\textrm{am}(u,\nu)\right),
\end{eqnarray}
where $\text{am}(u,\nu)$ is the inverse function of the incomplete elliptic integral of the first kind $F(\phi,\nu)$ for fixed $\nu$,
\begin{equation}
    u=F(\text{am}(u,\nu),\nu).
\end{equation}
In turn, the incomplete and complete elliptic integrals of the first kind are
\begin{eqnarray}\label{eq:incompleteelliptic}
\nonumber F(\phi,\nu)&\equiv&\int^\phi_0\frac{\mathrm{d}\varphi}{\sqrt{1-\nu\sin^2\varphi}},\\ K(\nu)&\equiv& F\left(\frac{\pi}{2},\nu\right).
\end{eqnarray}
As a consequence of these relations, $\text{sn}(u+2K(\nu),\nu)=-\text{sn}(u,\nu)$, $\text{cn}(u+2K(\nu),\nu)=-\text{cn}(u,\nu)$, and thus $\text{sn}(u,\nu),\,\text{cn}(u,\nu)$ are periodic functions with period $4K(\nu)$. 

The incomplete and complete elliptic integrals of the second kind are
\begin{eqnarray}\label{eq:incompleteellipticsecond}
E(\phi,\nu)&\equiv&\int_0^{\phi}\mathrm{d}\varphi~\sqrt{1-\nu\sin^2\varphi}\,,\\
\nonumber E(\nu)&\equiv& E\left(\frac{\pi}{2},\nu\right),
\end{eqnarray}
while those of the third kind are
\begin{eqnarray}\label{eq:incompleteellipticthird}
\nonumber \Pi(\phi,m,\nu)&\equiv&\int_0^\phi\frac{\mathrm{d}\varphi}{(1-m\sin^2\varphi)\sqrt{1-\nu\sin^2\varphi}},\\
\Pi(m,\nu)&\equiv&\Pi\left(\frac{\pi}{2},m,\nu\right).
\end{eqnarray}
Of interest are also the integrals
\begin{eqnarray}\label{eq:TechnicalIntegrals}
    F_{2n}(\nu)&\equiv&\int_0^{\frac{\pi}{2}}\frac{\sin^{2n}\phi}{\sqrt{1-\nu\sin^2\phi}}\mathrm{d}\phi,\\
    \nonumber  G_{2n}(\nu)&\equiv&\int_0^{\frac{\pi}{2}}\sin^{2n}\phi\sqrt{1-\nu\sin^2\phi}~\mathrm{d}\phi,
\end{eqnarray}
which appear when evaluating the particle number and energy of a cnoidal wave. They can be shown to obey the inter-recursive relations
\begin{eqnarray}
    F_{2n}(\nu)&=&\frac{F_{2n-2}(\nu)-G_{2n-2}(\nu)}{\nu},\\
    \nonumber  G_{2n}(\nu)&=&(2n-1)[G_{2n-2}(\nu)-G_{2n}(\nu)] \\
    \nonumber  &+&\nu [F_{2n+2}(\nu)-F_{2n}(\nu)],
\end{eqnarray}
yielding
\begin{equation}
    G_{2n}(\nu)=\frac{(2n-1)G_{2n-2}(\nu)+(1-\nu)F_{2n}(\nu)}{2n+1},
\end{equation}
with $F_0(\nu)=K(\nu),~G_0(\nu)=E(\nu)$.

Interestingly, by using elliptic functions, it is straightforward to solve analytically the textbook problem of the motion of a pendulum of mass $m$ subject to a constant gravitational field $g$:
\begin{equation}
    mL\ddot{\theta}=-mg\sin\theta,
\end{equation}
$\theta$ being the angle of separation with respect to the equilibrium position. Energy conservation reads
\begin{equation}
    \frac{1}{2}mL^2\dot{\theta}^2+mgL(1-\cos\theta)=E,
\end{equation}
which can be rewritten as
\begin{equation}
    \frac{1}{2}\dot{\theta}^2+2\frac{g}{L}\sin^2\frac{\theta}{2}=e\equiv \frac{E}{mL^2}.
\end{equation}
Integrating this equation yields
\begin{align}\label{eq:PendulumAnalytical}
\nonumber \theta(t)&=2\arcsin\left(\sqrt{\nu}~\textrm{sn}\left[\sqrt{\frac{g}{L}}(t-t_0),\nu\right]\right),~\nu\leq 1,\\
\theta(t)&=2\,\textrm{am}\left[\sqrt{\frac{e}{2}}(t-t_0),\frac{1}{\nu}\right],~\nu> 1,\\
\nonumber \nu&\equiv \frac{eL}{2g},
\end{align}
with $t_0$ a time-shift determined by the initial condition. The first line describes the bounded oscillation of a pendulum while the second line represents its full circular motion with an energy larger than that of the unstable equilibrium position ($\theta=\pi$). The corresponding periods of motion are
\begin{eqnarray}\label{eq:PendulumPeriod}
\nonumber T&=&4\sqrt{\frac{L}{g}}K(\nu),~\nu\leq 1,\\
T&=&\sqrt{\frac{8}{e}}K\left(\frac{1}{\nu}\right),~\nu>1.
\end{eqnarray}
The usual limit of small sinusoidal oscillation is recovered for $\nu\to 0$, where
\begin{equation}
    \theta(t)\simeq\sqrt{\frac{2eL}{g}}\sin\omega_0 (t-t_0),~\omega_0=\sqrt{\frac{g}{L}}.
\end{equation}

\subsection{GPG solutions in a ring}

Since the GPG equation can be rewritten as Eq. ($\ref{eq:GPGalileo}$), we first study in detail the solutions to the stationary homogeneous GP equation 
\begin{equation}
    H_{GP}\Psi_0=\mu\Psi_0.
\end{equation}
This equation is solved by decomposing the wavefunction in amplitude and phase as $\Psi_0(x)=A(x)e^{i\theta(x)}$, finding
\begin{eqnarray}
    -\frac{A''}{2}+\frac{v^2}{2}A+A^3&=&\mu A, \\
    \nonumber \frac{d\,n(x)v(x)}{dx}&=&0.
\end{eqnarray}
The second equation simply yields the conservation of the current, $J=n(x)v(x)$, with $n(x)=A^2(x)$ the density and $v(x)=\partial_x\theta$ the flow velocity. When plugged into the first equation, we obtain something of the form $A''=-W'(A)$, which can be thought as the equation of motion of a fictitious particle in a potential $W(A)$, where $x$ plays here the role of time. ``Energy'' conservation then implies:
\begin{eqnarray} \label{eq:GP-potential}
\frac{1}{2}A'^{2}+W(A) &=& E_A, \\
\nonumber W(A)&=&\frac{J^2}{2A^2}+\mu A^{2}-\frac{A^{4}}{2}.
\end{eqnarray}
By using instead the density, we find that the wavefunction is determined by the set of differential equations
\begin{eqnarray}\label{eq:ellipticdensities}
\nonumber n'^2&=&4(n-n_1)(n-n_2)(n-n_3),\\
\theta'(x)&=&v(x)=\frac{J}{n(x)},
\end{eqnarray}
where the three densities $0\leq n_1\leq n_2\leq n_3$ are the roots of the polynomial
\begin{equation}
     n^3-2\mu n^2+2E_A n-J^2=(n-n_1)(n-n_2)(n-n_3).
\end{equation}
These densities are related to the more physical chemical potential $\mu$, particle current $J$, and amplitude energy $E_A$ as
\begin{eqnarray}\label{eq:rootsumrule}
    \nonumber 
 \sum^3_{i=1}n_i&=&n_1+n_2+n_3=2\mu,\\
    \prod^3_{i=1}n_i&=&n_1n_2n_3=J^2,\\
    \nonumber n_1n_2+n_1n_3+n_2n_3&=&2E_A.
\end{eqnarray}
After integrating the equation for $n(x)$ in the first line of Eq. (\ref{eq:ellipticdensities}) and then that for $\theta(x)$ in the second line, we arrive at the cnoidal wave of Eq. (\ref{eq:CnoidalWave}). From its expression, we can compute the pure cnoidal contribution to the momentum 
\begin{equation}
    P_{\mathbf{n}}=\int^L_0-i\Psi^*_{\mathbf{n}}(x)\partial_x\Psi_{\mathbf{n}}(x)\mathrm{d}x=J L=\sqrt{n_1n_2n_3}L.
\end{equation}
Since the actual GPG wavefunction is $\Psi_0(x)=e^{ivx}\Psi_{\mathbf{n}}(x)$, it is immediately found that the total momentum is 
\begin{equation}\label{eq:CnoidalMomentum}
    P=Nv+P_{\mathbf{n}}\Longrightarrow v=\frac{\bar{p}}{\bar{n}}-\frac{\sqrt{n_1n_2n_3} }{\bar{n}}.
\end{equation}
Moreover, $|\Psi_0(x)|^2=|\Psi_{\mathbf{n}}(x)|^2$, and thus the total particle number is directly obtained from the integral of the cnoidal density. With the help of these results, one arrives at Eq. (\ref{eq:ClosedCnoidalSystem}). The trick to solve that system of equations is to express all roots $n_i$ in terms of $\nu$, $n_i=n_i(\nu)$. For that purpose, one combines the first two lines of Eq. (\ref{eq:ClosedCnoidalSystem}) to express $n_1,n_3$ in terms of $\nu$, and invokes the definition of $\nu$ itself, Eq. (\ref{eq:EllipticParameters}), to obtain $n_2(\nu)$. Inserting these results into the third line of Eq. (\ref{eq:ClosedCnoidalSystem}) yields a single equation for $\nu$, which can be easily numerically solved as $0\leq \nu \leq 1$. Once $\nu$ is obtained, the densities $n_i=n_i(\nu)$ are evaluated, and the complete GPG wavefunction and the total energy are determined. 


Finally, we show that the Berry-Gibbs curvature (\ref{eq:BerryGibbsCurvature}) automatically vanishes in our gauge choice. By decomposing the GPG wavefunction in phase and amplitude, $\Psi_0(\mathbf{x})=\sqrt{n(\mathbf{x})}e^{i\eta(\mathbf{x})}$, it is easily seen that in general
\begin{equation}\label{eq:BerryGibbsCnoidal}
    F_{AB}=\int\mathrm{d}\mathbf{x}~[\partial_A \eta\,\partial_B n-\partial_A n\,\partial_B \eta].
\end{equation}
In the specific case of the 1D GPG cnoidal wave, Eq. (\ref{eq:BlochWaveCnoidal}), $\eta(x)=(v+w_{\mathbf{n}})x+\Theta(x)$. Moreover, the second line of Eq. (\ref{eq:CnoidalMatching5}) imposes that the crystal momentum is fixed by the winding number and does not depend on the charges,
so $\partial_A\eta=\partial_A\Theta$. As in our particular gauge $n(x)$ is an even periodic function, then $\Theta(x)$ is an odd periodic function, and thus $F_{AB}$ identically vanishes by parity symmetry.

\bibliography{Hawking}
\bibliographystyle{quantum}

\end{document}